\documentclass[hidelinks,onefignum,onetabnum]{siamart171218}
\usepackage{subcaption,soul}
\usepackage{colortbl}
\usepackage{lipsum}
\usepackage{lineno}
\usepackage{amsfonts}
\usepackage{graphicx}
\usepackage{algorithmic}
\newcommand{\bt}[1]{\textcolor{black}{#1}} 
\definecolor{rred}{rgb}{0.,0.0,0.0}  
\newcommand{\red}[1]{\textcolor{rred}{#1}}

\usepackage{tabularx,booktabs}
\usepackage{multirow}
\usepackage{geometry}
\usepackage{xcolor} 

\newcommand{\Order}[1]{\ensuremath{\mathcal{O}(#1)}}    

\title{A projection method for particle resampling}
\author{Mark F. Adams\thanks{Lawrence Berkeley National Laboratory, Berkeley, CA (\email{mfadams@lbl.gov})}
\and Daniel S. Finn\thanks{National Research Council Naval Research Laboratory Postdoctoral Fellow}
\and Matthew G. Knepley\thanks{University of Buffalo, Buffalo, NY}
\and Joseph V. Pusztay\footnotemark[3]
}

\begin{document}

\maketitle

\begin{abstract}

Particle discretizations of partial differential equations are advantageous for high-dimensional kinetic models in phase-space due to their better scalability than continuum approaches with respect to dimension.
Complex processes collectively referred to as \textit{particle noise} hamper long time simulations with particle methods.
One approach to address this problem is particle mesh adaptivity, or remapping, known as \textit{particle resampling} \red{and \textit{remeshing}.}

This work introduces a resampling method that projects particles to and from a (finite element) function space.
\red{The method is simple, using standard sparse linear algebra and finite element techniques, and it preserves all moments up to the order of a polynomial represented exactly by the continuum function space.
It is distinguished from most other mesh-based methods in that new particle positions and number are decoupled from the mesh, allowing particle and continuum meshes to be adapted relatively independently. While this work is developed with structured particle and continuum phase-space grids on $1X$ + $1V$ Vlasov-Poisson models of Landau damping and two-stream instability, the method is well-suited to unstructured grids.}
\bt{Stable long time dynamics are demonstrated up to time $T=500$. Reproducibility artifacts and data are publicly available.}

\end{abstract}


\section{Background}
\label{sec:into}

Particle, marker particle, or macro-particle methods, such as particle-in-cell (PIC), are discretizations that, akin to traditional continuum-based methods such as finite elements (FE), finite volume, etc., discretize continuous PDE models as opposed to discrete, ground truth models, like molecular dynamics.
\red{Particle methods scale with dimension with a theoretical accuracy of $\Order{N^{-\frac{1}{2}}}$ and complexity $\Order{N}$, for $N$ particles. By comparison, continuum methods have higher order accuracy, $\Order{N^{-p}}$ for some order $p$, usually two or higher, and complexity $\Order{N^D}$, with $N$ grid points in each dimension $D$.}
For example, with a commonly attainable $p=2$ the complexities cross over at $D=4$ and particle methods have lower order complexity at higher dimensions.
This scaling with dimension has motivated the use of the PIC methods for the Vlasov-Maxwell-Landau (VML) system, or Boltzmann’s equation with Coulomb collisions in Landau form for magnetized plasmas and Vlasov-Poisson-Landau in astrophysics \cite{Lancellotti01112007}\red{. Continuum methods are a viable options as well \cite{Gkeyll,Shi2022}, and hybrid PIC / continuum methods have been developed \cite{KrausError2017,adams2024performance}.
In fact, the method developed here was motivated by conservative coupling of FE and PIC methods and results in a hybrid kinetic, continuum model.}  %

Mesh or grid adaptivity is a fundamental tool in PDE modeling for both continuum and particle methods, and is known as \textit{particle resampling} in particle methods.
This paper develops a particle resampling approach that uses a conservative mapping between particles and continuum grids, a projection \cite{PusztayKnepleyAdams2022}, that conserves an arbitrary number of moments exactly, that was developed by the structure preserving discretization community \cite{Kraus2017,WEINSTEIN1981235}.
This is flexible, allowing remapping to almost any new set of particles.
The focus of this paper is on investigating the efficacy of this method at reducing particle noise, adapting to improve grid quality, as well as undersirable side effects with standard plasma model problems.
The testing codes are built on PETSc (Portable, Extensible Toolkit for Scientific Computation), and are publicly available (Appendix \S\ref{sec:ad}).

Particle resampling methods have been developed by many groups.
Lapenta developed a method in the 1990s with a solve of the form $\left(M_pM_p^T\right)^{-1}$ with Lagrange multipliers that enforce moment constraints explicitly \cite{Lapenta1994213,Lapenta2002317,Lapenta1995139}, that is formally similar to the pseudoinverse solve in our approach \S\ref{sec:pseudo}.
Colella et. al., developed a \textit{particle remapping} method that is similar to our approach in a finite volume context with a \textit{direct remap} from the grid back to particles \cite{Colella2011, Myers2017, Myers16}.
\red{The Vortex-In-Cell community use resampling, or ``remeshing", extensively, for example, Morgenthal and Walther develop a method similar to Colella et. al. with high-order interpolation to a regular grid that preserves relevant moments and refers to much previous work in remeshing \cite{morgenthal2007a}.}
Faghihi et al. developed resampling methods with moment constraints and linear programming to enforce moments and other algebraic constraints \cite{FAGHIHI2020109317}.
Gonoskov proposed probabilistic down-sampling algorithms using algebraic constraints to enforced conservation \cite{GONOSKOV2022108200}.
Pfeiffer et al. introduced two conservative particle split and merge methods that use statistical properties of the plasma such as thermal speed \cite{PFEIFFER20159}.
Several groups have presented particle coalescence and splitting schemes, often using trees, within small groups of binned particles \cite{Teunissen2014318,Vranic201565,Assous2003550,Welch2007143,Luu2016165}.

Our approach is distinguished from previous works in a few ways.
An arbitrary number of moments are implicitly and provably conserved exactly, with exact linear solvers, without resorting to explicitly constraining the solution to desired manifolds.
Our method is simple and built entirely on standard sparse linear algebra primitives and allows for almost any new distribution of particles, the original particle grid being a good choice if dynamic adaptivity is not desired.
The Colella et. al. \red{algorithm uses phase-space adaptive mesh refinement to provide flexibility in particle positions and is similar to our algorithm if the resampling set is chosen to align with the vertices of the grid (this is explored in \S\ref{sec:directmap}) \cite{Myers16}, which removes the need for a pseudoinverse.}

This paper proceeds with relevant background in structure preserving methods in \S\ref{sec:sp}, the projection based resampling method in \S\ref{sec:pamr}, numerical methods and the test problems in \S\ref{sec:numr_method}, and \S\ref{sec:directmap} experiments with a direct remap method.
Experiments with the full high-order finite element projection and remapping method are presented in \S\ref{sec:results}\bt{, \S\ref{sec:longtime} investigates long time behavior with convergence studies, \S\ref{sec:problems} discusses side effects of resampling,} and \S\ref{sec:conc} concludes with a discussion of potential future work. 

\section{Structure preserving methods for Boltzmann's equations}
\label{sec:sp}

The critical idea in this work comes from research on structure preserving methods for Boltzmann's equations in general and the VML system for magnetized plasmas in particular.
Hamiltonian models in phase-space where density is a function of both space ($\mathbf{x}$) and velocity or 
momentum space ($\mathbf{v}$) are the fundamental equation of gravitational dynamics and electrostatics plasmas with the Vlasov-Poisson system, and electromagnetic plasmas with the Vlasov-Maxwell system.
A Coulomb collision term accounts for the statistics of particle interactions not present in the Hamiltonian \cite{hazeltine2013plasma,Hirvijoki2017,adams2024performance,Vocks_2002,2021APSDPPUP1078K}, giving rise to the governing equations for magnetized plasmas where the density of each species $\alpha$ is evolved in phase-space according to 
\begin{equation*}
\frac{df_\alpha}{dt} \equiv
\frac{\partial f_\alpha}{\partial t} + \frac{\partial  \vec{x}}{\partial t}
\cdot \nabla_x f_\alpha + \frac{\partial \vec{v}}{\partial t} \cdot \nabla_v f_\alpha = \sum_{\beta} C\left[f_\alpha,f_\beta\right]_{\alpha\beta},
\end{equation*}
where the collisional term is summed over all species $\beta$.
This equation is composed of the symplectic {\it Vlasov-Maxwell} term $\frac{df}{dt}=0$ and a metric, or diffusive, collision operator $C$.
Maxwell's equations provide an expression for $\frac{\partial \vec{v}}{\partial t} = a = \frac{q_\alpha}{m_\alpha}\left( \mathbf{E} + \mathbf{v} \times  \mathbf{B}\right)$.

This system has rich mathematical structure that can be preserved with proper discretizations.
The metriplectic formalism is an approach to analyze VML and to develop {\it structure preserving} discretizations \cite{Hirvijoki2017,Kraus2017,Hirvijoki2021}.
When a structure preserving grid-based collision operator \cite{Hirvijoki2017,AdamsHirvijokiKnepleyBrownIsaacMills2017, Adams2022a,adams2024performance} is coupled with a PIC method, a mechanism is needed to map distribution functions, in velocity space, between a particle representation and a finite element basis representation that preserves moments, as well as other structure \cite{PusztayKnepleyAdams2022}.
Preserving the second moment in velocity space, energy, is critical for many applications. 

\subsection{Structure preserving particle-finite element basis mapping}
\label{sec:sp2}

To apply a continuum operator in a PIC method that conserves moments a conservative particle-finite element basis mapping and remapping method is required.
Given a particle with weight $w_p$ and position $\textbf{x}_p$, a delta function representation $f_p(\textbf{x}) = w_p \delta(\textbf{x}-\textbf{x}_p)$, and a finite element (FE) space $V$ of functions $\phi_i$ and coefficients $\rho_i$, a function can be expressed as $f_{FE}(\textbf{x}) = \sum_i \rho_i \phi_i(\textbf{x})$.
Ideally $f_{FE}(\textbf{x}) = f_p(\textbf{x})$, but that is not possible.
Weak equivalence can however be enforced with: 

\begin{equation}
\int_\Omega d{\textbf{x}} ~\phi_j(\textbf{x}) \sum_p f_p(\textbf{x})=\int_\Omega d{\textbf{x}}~\phi_j(\textbf{x})\sum_p w_p \delta(\textbf{x}-\textbf{x}_p)=\int_\Omega d{\textbf{x}} ~\phi_j(\textbf{x})  f_{FE}(\textbf{x}) = \int_\Omega d{\textbf{x}} ~\phi_j(\textbf{x}) \sum_i \rho_i \phi_i(\textbf{x}) \forall \phi_j \in V. 
\label{eq:weak}
\end{equation}
With a \textit{particle mass matrix} $M_p[i,p] \equiv \phi_i(\mathbf{x_p})$, an FE mass matrix $M[i,j] \equiv \int_\Omega d{\textbf{x}} ~\phi_i(\textbf{x}) \phi_j(\textbf{x})$, a vector of particle weights $w$ and vector of FE weights $\rho$, (\ref{eq:weak}) can be written in matrix form as $$M \rho = M_p w,$$ which defines an equation for particle deposition on to the FE space:
\begin{equation}
\rho \leftarrow M^{-1}M_p w. \label{eq:dep}
\end{equation}
This mapping is proven to conserve moments up to the order polynomial that the FE space can represent exactly \cite{KrausError2017,PusztayKnepleyAdams2022}, for example a quadratic element mesh in velocity space is sufficient to conserve energy.

After deposition on the FE space, a Poisson or Ampere's law solve can be applied or a collision operator, $L$, can be evolved, $u \leftarrow L \rho$.
In mapping $u$ back to particles one can simply invert (\ref{eq:dep}), $\overline w \leftarrow M_p^{-1} M u$, however $M_p$ is rectangular in general.
The key idea is that a pseudoinverse, $M_p^{\dagger}$ with $M_p M_p^{\dagger} = I$, conserves moments with standard sparse linear algebra: 
\begin{equation}
\overline w \leftarrow M_p^{\dagger} M u.
\label{eq:remap}
\end{equation}
Thus, if the collision operator conserves moments this entire process of applying a continuum operator in a PIC method conserves moments \cite{Hirvijoki2017}.

\subsection{Pseudoinverses and idempotent projections}
\label{sec:pseudo}

There are two basic approaches to the pseudoinverse: an appropriate Krylov methods such as LSQR or Moore-Penrose.
Both of these solvers are $l_2$ projections, but there are alternative norms, such as $L_2$, that could be investigated.
Moore-Penrose is attractive because it is easier to precondition a square matrix, especially for batch solvers \cite{adams2024performance}.
Krylov methods are attractive as they can solve singular systems transparently.
Preconditioning LSQR requires some effort, but the pseudoinverse solves in this work are very well conditioned and unpreconditioned LSQR works well.
Though we use LSQR in this work, the analysis is clearer with the Moore-Penrose pseudoinverse: $M_p^{\dagger} \equiv M_p^T \left(M_P M_p^T \right)^{-1}$, and stability is easier to understand. 

If a collision operator is not used, $L=I$, then combining (\ref{eq:dep}) and (\ref{eq:remap}) results in the remapping algorithm 
\begin{equation}
\overline w \leftarrow M_p^T \left(M_P M_p^T \right)^{-1}M_p w,
\label{eq:cg}
\end{equation}
which is a type of ``coarse-graining" algorithm, a mechanism to add numerical entropy dissipation developed in the physics community \cite{Chen2007-CG,Vernay2012}.

\paragraph{Idempotent property of projections}
Information is lost when projecting a particle representation of a function onto a FE basis if the number of particles exceeds the number of FE basis functions, which is the case of interest.
An attractive property of (\ref{eq:cg}) is that information is only lost on the first application of coarse graining in that $$ \overline{\overline{w}} = M_p^T \left(M_P M_p^T \right)^{-1}M_p {\overline w} = M_p^T \left(M_P M_p^T \right)^{-1}M_p M_p^T \left(M_P M_p^T \right)^{-1}M_p w = M_p^T \left(M_P M_p^T \right)^{-1}M_p w = {\overline w}.$$
Thus $\overline{\overline{w}} = {\overline w}$ and the coarse-graining operator is idempotent, which is an elegant property in that this process does not, in a sense, evolve the operator although it does add diffusion.

\section{A particle resampling method}
\label{sec:pamr}

A key observation is that after computing (\ref{eq:dep}), the distribution function is entirely represented on the FE space and particle weights and positions are no longer needed.
A new set of particle positions, essentially any new set, can be created.
A new particle mass matrix, $\overline M_p$, can then be computed and (\ref{eq:cg}) can be continued with $\overline M_p$.
Moments are conserved because it is provable, and experimentally demonstrated, that the projection to the grid preserves moments and the projection from the grid preserves moments.
This resampling method rearranges (\ref{eq:cg}) by projecting back to a new set of particles after the deposition according to:
\begin{itemize}
    \item use (\ref{eq:dep}) to deposit the distribution function on to the FE grid $c\leftarrow M_p w$,
    \item create a new set of particles to generate a new particle mass matrix $\overline M_p$,
    \item apply a pseudoinverse to compute weights for the new particles $\overline w \leftarrow \overline M_p^{\dagger} c$.
\end{itemize}
\vspace{3mm}
\paragraph{Field preservation with resampling: ${\overline \rho} = \rho$} An attractive property of this resampling is that the right hand side of the field solves, Poisson and Ampere's law solves, or collision operators are not affected by the resampling:

$$ \overline{\rho} = M^{-1}{\overline M_p} \overline{w} = M^{-1}{\overline M_p} {\overline M_p}^T \left({\overline M_p} {\overline M_p}^T \right)^{-1}M_p w =  M^{-1} M_p w = \rho. $$

\subsection{Moore-Penrose stability}
\label{sec:mp}

Care must be taken in the explicit inverse of $M_p M_p^T$ as it can be singular.
Stability can be guaranteed because defining the particle grid is under the control of the algorithm, unlike in coarse-graining and field solves.
A necessary condition for stability of the pseudoinverses is that there does not exist a set of vertices whose union of support of associated FE basis functions contains less particles than the number of vertices in the set.
This is not rigorous but comes from the intuition that no set of equations (rows of $M_p$) should have less non-empty columns than rows, otherwise the matrix locally singular.
Further we find that more particles than basis functions are required for stability in, for example, the $1D$ periodic direction if, with degree $Q$ element (e.g., $Q=2$, a Q2 element), there are $Q$ particles per cell, which results in an equal number of particles and basis functions (equations or vertices).
$M_p$ is square in this case, but we observe that $\left(M_P M_p^T \right)$ is singular except for the special case where the particles and mesh points are aligned as in the direct remap method in \S\ref{sec:directmap}.
This criterion is not practical to check with an arbitrary set of particles, and we use $\left(Q+1\right)^{D}$ particles per cell, which is safe and simple, but one could probably create a grid with $Q^D + 1$ particles in general positions.

{\color{black}
\subsection{Complexity}
\label{sec:complex}

The addition of this resampling method in a PIC code results in a hybrid of particle and continuum methods, similar to the effect of using a grid based collision operator in a PIC code \cite{KrausError2017,adams2024performance}.
As discussed in \S\ref{sec:into} with a simple complexity analysis, continuum methods do not scale well with dimension and a more detailed complexity analysis of this method is warranted.
In a kinetic plasma model, the resolution requirements in velocity space are much smaller than in configuration space. 
Meshes are application dependent, but velocity space distributions do not have the potential for the geometric complexity of configuration space in, say, a fusion device or solar coronal jets.
There are no boundaries or material properties (e.g., engineered structures like walls) in velocity space and distributions tend toward smooth near-Maxwellians that can be resolved with about $10^3$ mesh points \cite{adams2024performance}, however sources such as antennae for RF heating \cite{Yajima_2019}, alpha particles from fusion processes \cite{Hu_2025} and runaway electrons \cite{Adams2022a}, can result in localized structure that may require, say, $10^4-10^5$ mesh points.
A well resolved tokamak model uses about $10^8$ points in configuration space.
Thus, one can think of $6D$ kinetic models of magnetized plasmas as scaling roughly like a $4.5D$ model (and more like $4D$ models with a gyrokinetic approximation).
Note, this observation indicates why continuum methods are feasible for these problems. 

The work and memory complexity of the pseudoinverse is composed of the work of building the particle mass matrix and the pseudoinverse solve for each species. 
For analysis, assume tensor product Cartesian grids of, in the $6D$ case, two $3D$ grids of the same size for simplicity of analysis with tensor elements, and a constant number of particles per phase-space cell.
A particle is interpolated to three vertices in each dimension with $Q2$ elements, which results in $3^6 = 729$ non-zeros per row of $M_p$ in $6D$.
This is substantial and suggests the use of a matrix-free $M_p$ solver, which is sufficient for the unpreconditioned LSQR solver used in this work.
The solves in the test problems converge to high accuracy, to conserve moments, in about 50 iterations.
\paragraph{Memory complexity}
Stability of the pseudoinverse requires there be more particles than vertices.
To ensure this reliably, assume $Q+1$ particles per cell in each dimension ($Q$ is the lower bound for stability).
There are $NQ$ vertices in each dimension, asymptotically with $N$ cells, which results in a ratio of number of particles to number of vertices of $\left(\frac{Q+1}{Q}\right)^{6} \approx 11$.
The lower bound is one.
Particle methods use at least $D+1$ words per particle and one word for each species each mesh vertex for number density.
The $7-8$ work vectors in LSQR would need to be explicitly stored, resulting in roughly the same memory complexity for both grids and particles, depending on the number of particles per vertex and the number of species, which is assumed to be one herein.
\paragraph{Particle work complexity}
The work complexity of applying $M_p$ is composed of: 1) \Order{D(Q+1)^{D+1}} flops per cell for the spectral element Jacobian with sum factorization that is amortized with the $(Q+1)^D$ particles per cell, 2) \Order{D (Q+1)} flops per particle to evaluating shape functions, and 3) \Order{(Q+1)^D} flops in applying the interpolant to the source vector for each particle.
The application of the interpolants is the dominant cost, which is of the same order as the memory complexity that we avoid with matrix-free.
This complexity could probably be improved with optimized algorithms.
\paragraph{Particle data movement complexity}
The data movement complexity per cell includes 1) reading $D(Q+1)^D$ words of cell closure coordinates to compute the element Jacobian, 2) reading $D$ words of particle coordinate data or, with $(Q+1)^D$ particles per cell, $D(Q+1)^D$ words per cell, to compute the interpolants, 3) reading $(Q+1)^D$ words from the source vector and 4) read/write $(Q+1)^D$ words from/to the destination vector per cell.
Again, this naive result could no doubt be improved with algorithmic development in the future.

}

\section{{Numerical tests and the PETSc test harness}}
\label{sec:numr_method}

The testing code for this paper is built on the PETSc-PIC framework~\cite{Finn2023,pusztay2023landau,PusztayKnepleyAdams2022}, a recently developed PIC toolkit in the PETSc~\cite{petsc-web-page,PETScManual}. The PETSc-PIC framework primarily relies on two modules to drive forward the particle and finite element space. These modules are \texttt{DMSwarm}~\cite{MayKnepley2017} and \texttt{DMPLEX}~\cite{LangeMitchellKnepleyGorman2015}, respectively. \texttt{DMSwarm} provides a fully parallel solution for particle methods and for particle-mesh methods while \texttt{DMPLEX} provides generic unstructured mesh creation, manipulation and I/O. 

The finite element method (FEM) is used to solve the field equations at each timestep. The PETSc FEM framework abstracts the construction of the finite element using the Ciarlet triple~\cite{Ciarlet1976}, consisting of a mesh object (\texttt{DMPLEX}), a finite-dimensional function space (\texttt{PetscSpace}), and a dual space (\texttt{PetscDualSpace}). This is all handled by the \texttt{PetscFE} object and can be customized from the command line. In previous work~\cite{Finn2023}, simple $H^1$ finite element spaces have been sufficient in capturing the short timescale linear plasma kinetics. Thus, we will continue the use of these $H^1$ spaces in this work.

Particle pushing for the VP system relies on the characteristics of linear hyperbolic Vlasov equation which may be derived by first writing a simplified form of the Vlasov equation,
\begin{gather}\label{advection}
\frac{\partial f_\alpha}{\partial t} + \frac{\partial  \mathbf{x}}{\partial t}
\cdot \nabla_x f_\alpha + \frac{\partial \mathbf{v}}{\partial t} \cdot \nabla_v f_\alpha = 0 \\
\frac{\partial f_\alpha}{\partial t}+\mathbf{z} \cdot \nabla_{\mathbf{q}} f_\alpha = 0,\nonumber
\end{gather}
where $\mathbf{q} = (\mathbf{x}, \mathbf{v})$ is the phase-space variable and $\mathbf{z} = (\mathbf{v}, -q_e \mathbf{E}/m)$ is the combined force. The force term $-q_e \mathbf{E}/m$ is independent of velocity, and therefore \eqref{advection} may be written in the conservative form,
\begin{equation}
  \label{cons advection}
  \frac{\partial f_\alpha}{\partial t}+  \nabla_{\mathbf{q}} \cdot (\mathbf{z}f_\alpha)=0.
\end{equation}
Given this advective form of the Vlasov equation, we can rewrite the equation for the characteristics $\mathbf{Q} = (\mathbf{X}, \mathbf{V})$,
\begin{equation}
  \frac{d \mathbf{Q}}{d t} = \mathbf{z},
\end{equation}
which re-expressed with the original phase-space variables gives,
\begin{align}
  \frac{d \mathbf{X}}{d t} &= \mathbf{V}, \\
  \frac{d \mathbf{V}}{d t} &= -\frac{q_e}{m} \mathbf{E}. \nonumber
\end{align}
Since particles follow characteristics, the Vlasov equation in the particle basis becomes
\begin{align}
  \frac{d \mathbf{x}_p}{d t} &= \mathbf{v}_p, \\
  \frac{d \mathbf{v}_p}{d t} &= -\frac{q_e}{m} \mathbf{E}. \nonumber
\end{align}

Solving the characteristic equations is conducted in PETSc with the \texttt{TS} module using explicit symplectic integrators, a subclass of geometric integrators introduced by Ruth in~\cite{Ruth1983}. 
In general, for PIC models, explicit integrators are not energy conservative and have a tendency to increase total energy over long time scales through ``numerical heating''. In previous work~\cite{Kraus2017,Finn2023,pusztay2023landau}, however, explicit symplectic integrators have been shown to achieve exact conservation of mass and momentum, as well as a stable approximate conservation of the system energy. The PETSc \texttt{TS} module contains well-tested implementations for first- to fourth-order symplectic integrators. For full Tokamak models and collisional cases, the \texttt{TS} module also contains a variety of implicit time integrators. These include the recently added discrete gradients method which has been tested on both VP and collisional Landau systems~\cite{Finn2025}. With these implicit methods, larger time steps can be taken while remaining stable, capturing long time physical phenomena. Furthermore, exact energy conservation has been previously shown using implicit methods~\cite{Markidis2011}. In this work, however, we are interested in capturing the fastest waves in the Landau damping system. Thus, explicit methods are more appropriate and less costly than implicit integrators. We choose a first-order symplectic integrator, \textit{symplectic Euler}.

\subsection{\red{Two-stream instability}}\label{sec:TwoStreamTheory}

{\color{black}
In this work, we first study the one-dimensional two-stream instability test, often used as a plasma benchmark~\cite{BirdsallLangdon}. This test is characterized by an exponential growth in the electric field, meaning the buildup of statistical noise has less impact on the system than other plasma tests. This makes it a useful control for a resampling algorithm. 
To focus on the electrostatic kinetic effects, we ignore collisional dynamics and reduce the full VML equations to the collisionless, magnetic-free Vlasov-Poisson (VP) system.

The initial system consists of two counter-streaming thermal electron populations with a small perturbation in physical space, described by the distribution,
\begin{gather}
    f(x,v,t=0) = \frac{1}{2\sqrt{2\pi v_{th}^2}} \left( e^{-(v - v_d)^2/2v_{th}^2} + e^{-(v + v_d)^2/2v_{th}^2} \right) \left(1+A \cos\left(k x\right)\right),\\
    (x,v) = \left[0,4\pi/k\right]\times\left[-v_{max},v_{max}\right],
\end{gather}
where $k=0.5$ and $A=0.01$ are the initial perturbation wavenumber and amplitude, $v_d=2.0$ is the drift velocity, i.e., the beam spread, $v_{th}=0.5$ is the thermal velocity of the plasma and $v_{max}=10$. 
In the test, the plasma wave generated by the particles interacts with each counter-streaming population causing particles moving slower than the wave to be accelerated and particles moving faster than the wave to be decelerated. The net deceleration of particles from the two populations leads to a period of linear growth in the electric field strength which becomes unstable due to the initial fluctuations introduced in the distribution function. After the period of electric field growth, resonant particles ($v \approx v_{th}$) become trapped by oscillatory orbits inside potential wells leading to a saturation and flattening in the electric field strength and formation of phase-space vortices. 

The evolution of small linear perturbations in the plasma can be described by the dispersion relation,
\begin{equation}
    1 + \sum_s \frac{\omega_{p,s}^2}{k^2} \int \frac{\partial f_0/\partial v}{v-\omega/k} \mathrm{d}v = 0,
\end{equation}
which arises from the linearization of the Vlasov–Poisson system. If the slope of the distribution function between the beams is positive at the field's phase velocity, $v_{\phi}=\omega/k$, i.e., $\partial f_0/\partial v > 0$, the instability appears. From this relation, we can calculate the linear growth rate of the electric field, $\gamma=Im(\omega)$, as our figure of merit in the verification tests. The later stage phenomena, e.g., electric field saturation and vortex formation, become nonlinear in nature and are no longer described by the dispersion relation. Thus, they are not used for quantitative verification in the test. However, they are useful in observing the qualitative effects of the resampling methods on the grid and particles, as will be discussed in later sections.

The linear growth phase is marked in Figure~\ref{fig:TS_ef1} by the red dotted fit line. We consider the period after the fit line ends to be the beginning of the nonlinear phase. It is near this boundary between linear and nonlinear phases that the second and third phase-space plots in Figure~\ref{fig:TS_pp1} are generated.
}




\subsection{Landau damping}

\red{As a more thorough test of noise reduction with resampling, we} consider the classic one-dimensional plasma test, Landau damping, given by the initial state,
\begin{gather}
    f(x,v,t=0) = \frac{1}{\sqrt{2\pi}}e^{-v^2} \left(1+A \cos\left(k x\right)\right),\\
    (x,v) = \left[0,2\pi/k\right]\times\left[-v_{max},v_{max}\right],
\end{gather}
where $k=0.5$, $v_{max}=6$, and we consider three values for the wave amplitude, $A$: $0.5$, $0.01$ and $0.0001$. \red{In this test, we have normalized the thermal velocity, $v_{th}$, to $1$}. The Landau damping test is a popular choice for Vlasov benchmarking because it involves a number of purely kinetic effects, such as phase mixing, and it has simple analytical solutions. A detailed description of the analytical solutions to the Landau damping problem can be found in~\cite{Finn2023}. \red{The key difference between this test and two-stream instability is the now negative slope of the distribution, $\partial f_0/\partial v<0$, at the wave phase velocity, $v_{\phi}$, which leads to the damping effect observed in the field.} 

The two cases where the wave amplitude, $A$, is $0.01$ and $0.5$ are referred to as the \textit{linear} and \textit{nonlinear} landau damping test cases, respectively~\cite{KrausThesis,Cheng1976}. A third test case, $A=0.0001$, sits well within the linear regime and is also common. In the linear case, the small field perturbation is damped out at a rate of $\gamma=-0.153$ in favor of a more homogeneous field. However, in collisionless tests the growth of subgrid modes can disrupt the field damping and cause large gradients to develop in the phase-space. These large gradients lead to a sudden regrowth of the field. Previous work~\cite{KrausThesis} has show that the inclusion of collisions can remove these subgrid modes, damping the field to machine precision with a continuum code.

In the nonlinear case, $A=0.5$, the field decay reverses much earlier and the general dynamics differ from that of the linear case. The two primary explanations for this earlier field resurgence are the stronger interaction between the potential well created by the electric field which resonates with and accelerates more of the particles, and the increased phase mixing, evident in phase-space diagrams. To fully capture these dynamics, the nonlinearized form of the Vlasov equation must be considered. Significant work was done by Villani and Mouhot in~\cite{Villani2009} to analyze the nonlinear Vlasov equation and show that, while the nonlinear dynamics present in this system lead to a weaker initial decay of the electric field, over long enough time scales, the field will damp out, as it does in the linear case. Thus, it is vital to develop the tools necessary to capture the long time evolution of these kinetic plasmas structures.
From~\cite{KrausThesis} and~\cite{Cheng1976}, we expect the initial damping rate of the field to be $\gamma_1=-0.286$ which quickly turns into a field growth at a rate of $\gamma_2=0.086$. We will use these values to verify our tests in later sections.


\subsubsection{Continuum grids}
\label{sec:part_grids}

The phase-space continuum grids in 
\red{the implementation} are on regular 90 degree lattice with an option for simple r-refinement in velocity space.
Cartesian sub-particle grids are defined in each phase-space cell, similar to Lapenta (Figure~1, \cite{Lapenta1995139}).
For simplicity, the original grid is used for resampling in this \red{implementation} and adaptivity strategies are left for future work.
As discussed in \S\ref{sec:mp}, with periodic boundary conditions in the spatial dimension and natural boundary conditions in velocity dimension we use at least $Q+1$ particles in each dimension in each phase-space cell for stability of the pseudoinverse. 

\begin{figure}[h!]
    \centering
    \includegraphics[width=0.39\linewidth]{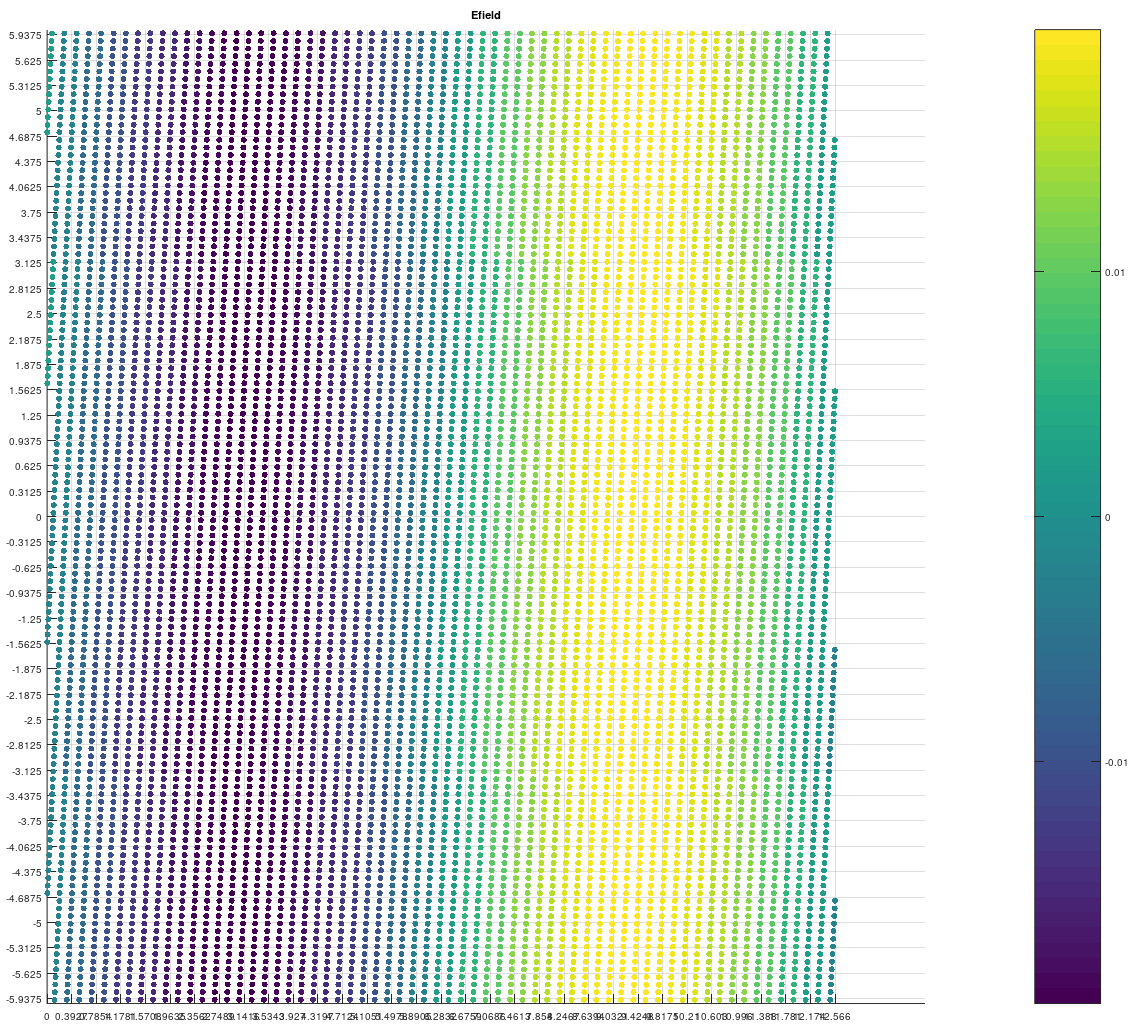}
    \includegraphics[width=0.39\linewidth]{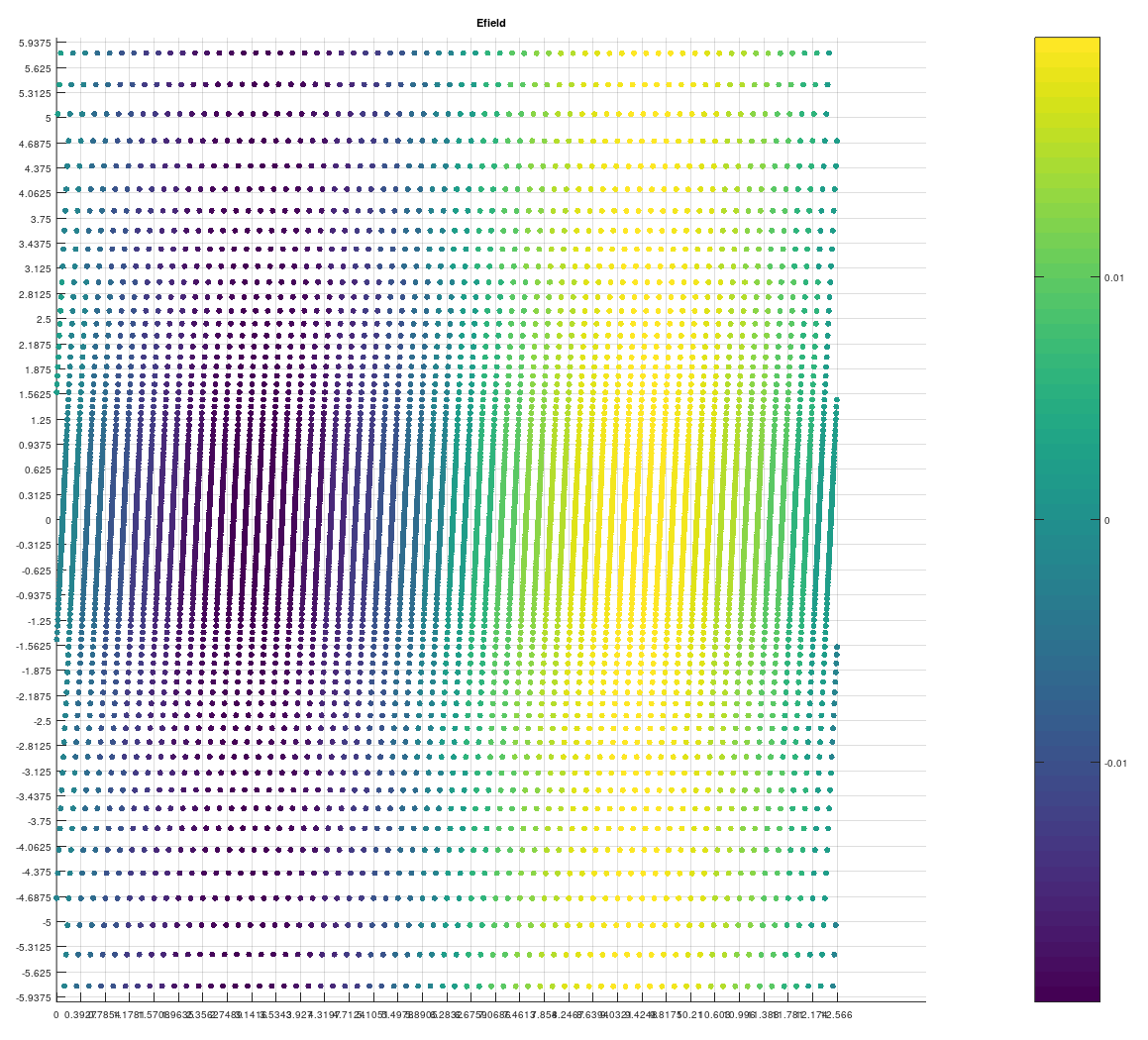}
y-   \caption{E field on $64 \times 128$ particle grid (y-axis is velocity): uniform distribution (left); r-refinement (right)}
    \label{fig:rref}
\end{figure}

The test harness is equipped with a simple \textit{r-adaptivity} capability where points are pushed toward the origin in velocity space to better represent a Maxwellian distribution.
Figure~\ref{fig:rref} shows the electric field (E) on uniform and r-refined versions of a $64 \times 128$ particle grid $X \times V$, with particle clustering around $v=0$ and the initial perturbation in $x$ of the electric field.
\bt{Data with r-refinement grids have ``graded" in the title of the plot and uniform grid data have ``uniform" in the title.
Similarly, the order of the finite element space in the data is encoded in the title, for example, $Q_x1-Q_v2$ uses linear $Q1$ (or $P1$) elements in real space and quadratic elements in velocity space.}




\section{\bt{Direct remap method with low-order finite elements}}
\label{sec:directmap}
{\color{black}
Colella et al. developed a method that can be viewed as a version of our algorithm in a finite volume context. 
In this algorithm, the continuum grid data after particle deposition, is mapped directly to new particles \cite{Colella2011,Myers2017,Myers16}.
Cubic splines are used for particle deposition on Cartesian grids.
They demonstrate high-order convergence \cite{Myers2017}, and use phase-space adaptive mesh refinement (AMR) grids \cite{Myers16}.
This method should conserve energy because the cubic spines can represent a second order polynomials exactly and the direct remap obviously conserves all moments in projecting the grid to particles.

The test harness supports a direct map method in a vertex-centered, finite element context.
The grid vertices define the new particles and the domain is doubly periodic (to make $M_p$ square), which is immaterial given that density is negligible at the velocity boundary.
The salient feature of this construct is that the particle mass matrix is the identity, $\overline M_p = I$, and the pseudoinverse vanishes.

\begin{figure}[h!]
    \centering
    \includegraphics[width=.68\linewidth]{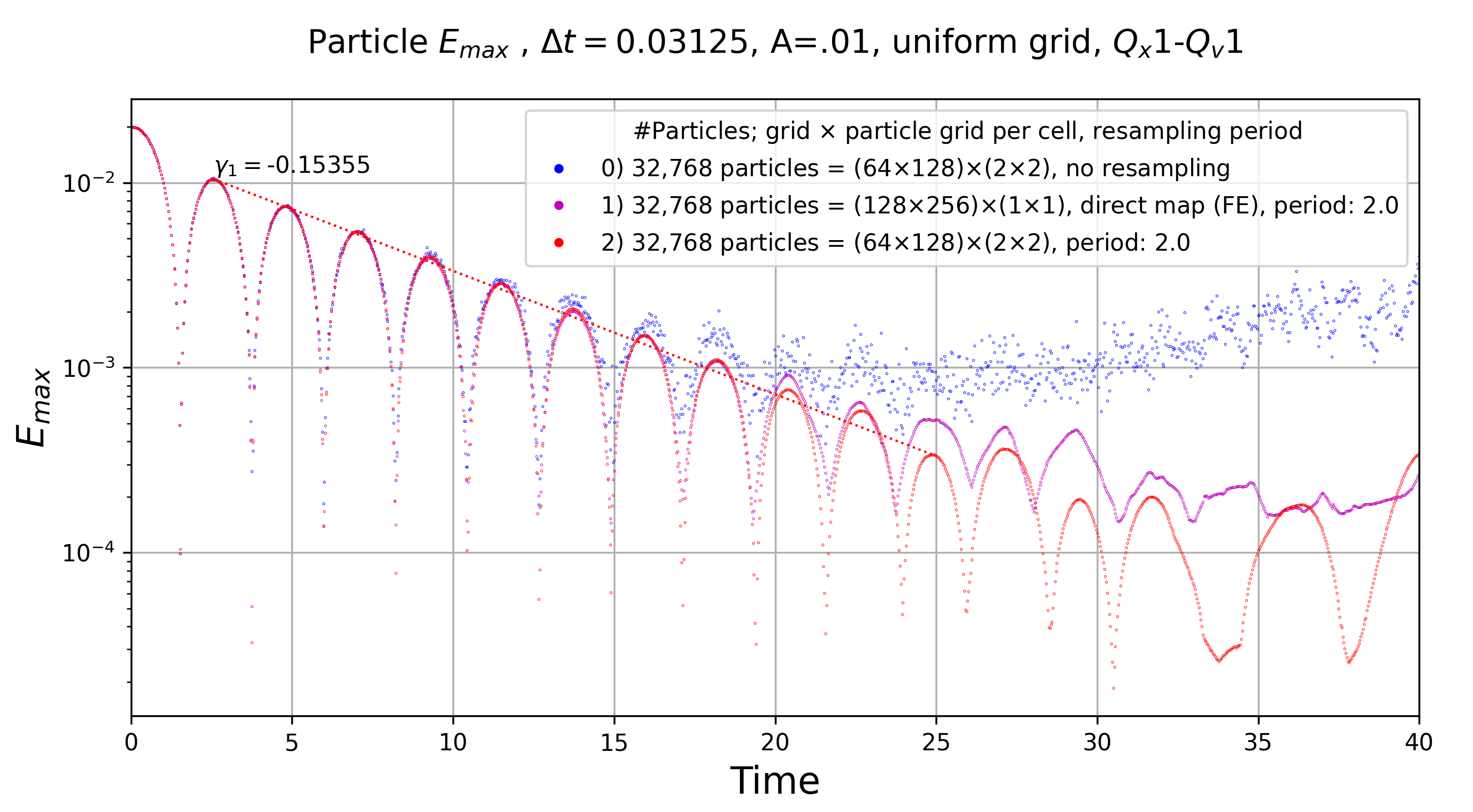}
    \caption{Field amplitude with linear Landau damping, $A=0.01$: no resampling (blue and noisy), a direct remapping finite element version of Myers et al. (magenta), and the projection method (red)}
    \label{fig:myers}
\end{figure}
Figure~\ref{fig:myers} show the results of this direct map algorithm with the psuedoinverse project on a linear Landau damping, $A=0.01$, problem with a $128 \times 256$ particle grid and $V_{max} = 6.0$ and results in agree well with amplitude reported in \cite{Myers2017} (Figure~3.1).
$Q1$ elements are used for all spaces.
This direct map method requires one particle per cell and the projection method requires two particles per cell in each dimension to ensure stability.
To maintain the same number of particles for all three tests, the projection method uses a $2\times 2$ particle grid per cell and half as many cells in each dimension.
This data shows again, but with $A=0.01$, the efficacy of resampling in that it suppresses the noise observed without resampling.
}

\section{Numerical experiments with pseudoinverse resampling}
\label{sec:results}

This section investigates this projection resampling method on a \red{two-stream instability, a linear Landau damping and a nonlinear Landau damping test}.
One issue to be addressed in a PIC method is ensuring a $C^0$ electric field, which our $C^0$ discretization of the Poisson equation does not provide given that one order of continuity is lost in the gradient of the potential.
One can use an $H(div)$ Poisson solver or $C^1$ finite elements, but with $C^0$ elements we project the field to the vertices and then back to the particles, resulting a $C^0$ electric field.
This method helps to stabilize our method, and is all but required for $Q2$ elements in space.
These instabilities could potentially be addressed with a quite start method \cite{SYDORA1999243}.
Mesh adaptivity in velocity space \cite{adams2024performance}, may act as a type of quite start, and is a subject of future work.


\subsection{{Two-stream instability}}

{\color{rred}

For the two-stream instability test, we consider a $60\times400$ particle grid with a $2 \times 3$ particle grid per cell. This gives a total particle number of $144,000$. Particle and mesh convergence studies were used to calculate this ideal grid which achieves low error and does not require significant computational costs. As discussed in~\S\ref{sec:part_grids}, $Q1$ elements are used for real space and $Q2$ elements for velocity space. 

We consider the results of the two-stream instability test both quantitatively and qualitatively. For the quantitative analysis, we compare the linear growth rate of the electric field for a run without resampling and several runs with different resampling periods to the analytic growth rate. Given the initial distribution defined in~\S\ref{sec:TwoStreamTheory}, we expect this analytic growth rate of the electric field to be $\gamma=0.20488$. Figure~\ref{fig:TS_ef1} shows the evolution of the max norm of the electric field, with and without resampling. The linear growth phase is marked in Figure~\ref{fig:TS_ef1} with a dashed red line.  Table~\ref{fig:ts_errors} shows the calculated growth rate and the error in the fit for each run. The data shows a general trend of decreasing error when more resampling steps are taken. However, the improvement in accuracy is small. During each resampling step, we also calculate the second moment, kinetic energy, of the plasma before and after the step. In each of the runs, the change in the kinetic energy during the resampling step is $\mathcal{O} (10^{-12})$, which falls within solver tolerances. Thus, the algorithm is conservative when using $Q2$ elements for velocity space.
\begin{figure}[h!]
    \centering
    \begin{subfigure}[b]{1\textwidth}
       \centering
       \includegraphics[width=1\linewidth]{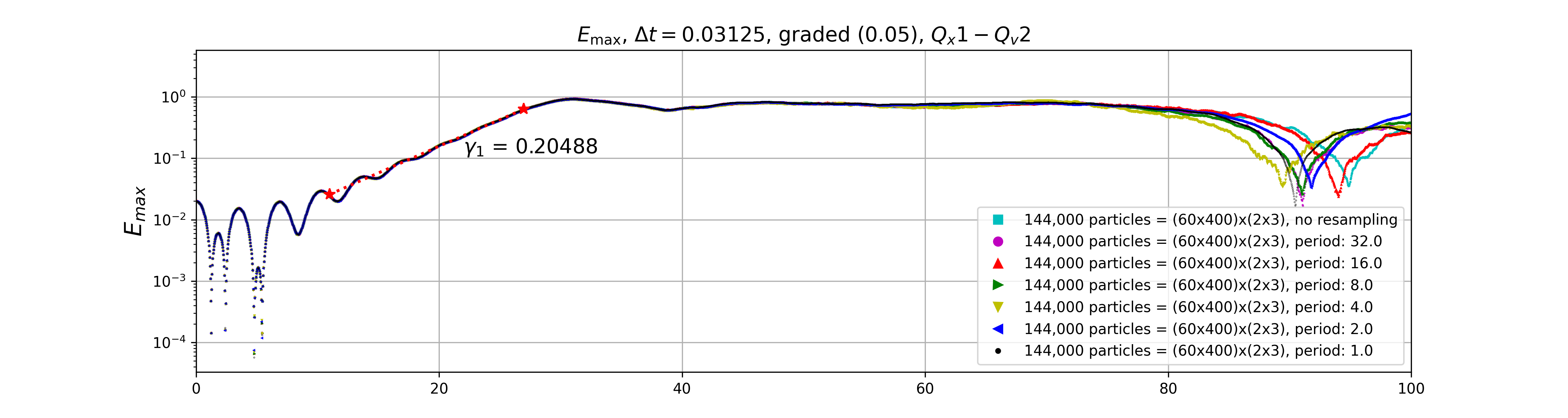}
       \caption{Linear growth phase and early nonlinear phase.}
       \label{fig:TS_ef11}
    \end{subfigure}\\%
    \begin{subfigure}[b]{1\textwidth}
       \centering
       \includegraphics[width=0.85\linewidth]{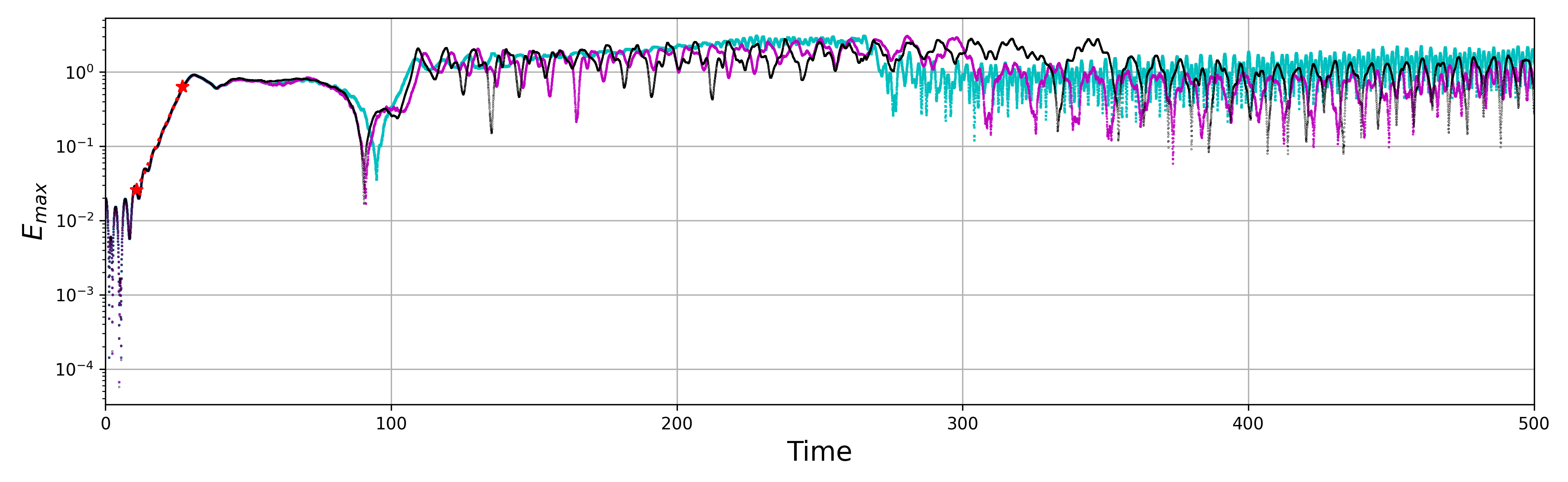}
       \caption{Entire runtime.}
       \label{fig:TS_ef12}
    \end{subfigure}
    \caption{$E_{max}$, Two-stream instability resampling period comparison test.}
    \label{fig:TS_ef1} 
\end{figure}
\begin{table}
    \centering
    \begin{tabular}{|c|c|c|}
        \hline
        Resampling Period & $\gamma_{fit}$ & $\gamma_{error}\;\left[\%\right]$ \\
        \hline\hline
        1.0 & 0.20013 &  2.31684 \\
        2.0 & 0.19999 &  2.38579 \\
        4.0 & 0.19993 &  2.41687 \\
        8.0 & 0.19993 &  2.41358 \\
        16.0 & 0.19985 & 2.45319 \\
        32.0 & 0.19987 & 2.44742 \\
        No Resampling ($\infty$)& 0.19987 &  2.44742 \\
        \hline
    \end{tabular}
    \caption{Growth rate errors in two-stream instability resampling rate comparison test.}
    \label{fig:ts_errors}
\end{table}

Qualitatively, we consider the effect resampling has on preserving the grid of particles in the system and on the electric field growth in a long-time simulation. Between $T \sim 90$ and $T \sim 110$, the two ``stable'' phase-space vortices begin to interact and merge into a single vortex, as seen more clearly in Figure~\ref{fig:TS_pp1}. Furthermore, after $T = 300$, these vortices disappear altogether, replaced by repeated unstable vortex formations and collapses. These periods of change are also noted in Figure~\ref{fig:TS_ef1}, where the electric field briefly declines and then proceeds to grows back stronger. The merging of the vortices causes a breakdown of the coherent charge separation that sustains the electrostatic wave, leading to the brief conversion of field energy back to kinetic energy. The inclusion of resampling appears to shift this vortex merging to earlier times in the simulation. It is likely that the resampling algorithm smooths phase-space filaments and disturbs the relative phase between the particles and wave causing the earlier breakdown of the stable vortices~\cite{Bernstein1957}. This breakdown of the stable vortices is also accompanied by an increase of the kinetic entropy of the system. However, this effect is not measured here and will be the focus of future work.
\begin{figure}[h!]
    \centering
    \includegraphics[width=1\linewidth]{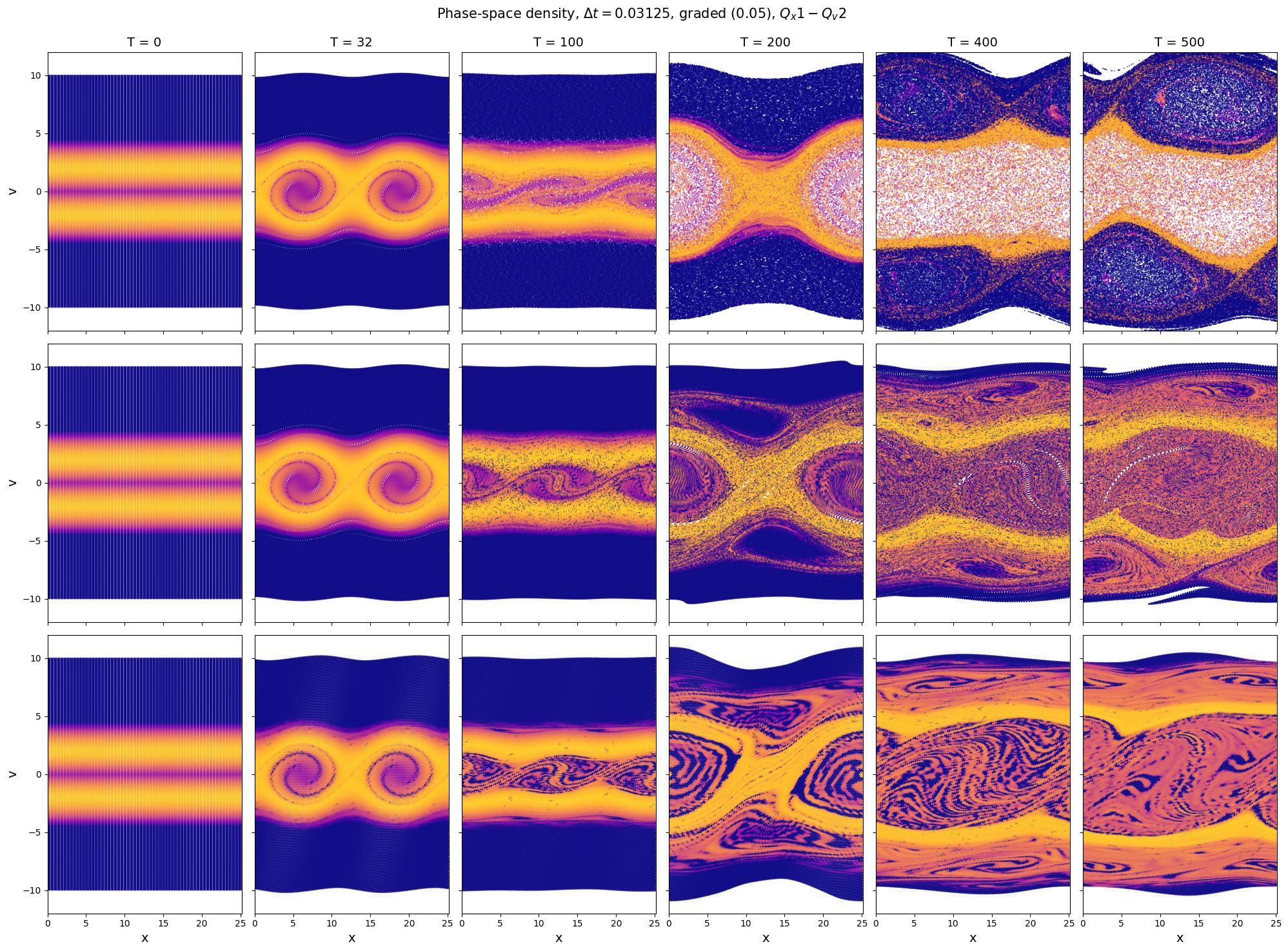}
    \caption{Two-stream instability phase-space density evolution for three cases: (top row) no resampling, (middle row) resampling period $32.0$ and (bottom row)  resampling period $1.0$.}
    \label{fig:TS_pp1}
\end{figure}
\begin{figure}[h!]
    \centering
    \includegraphics[width=1\linewidth]{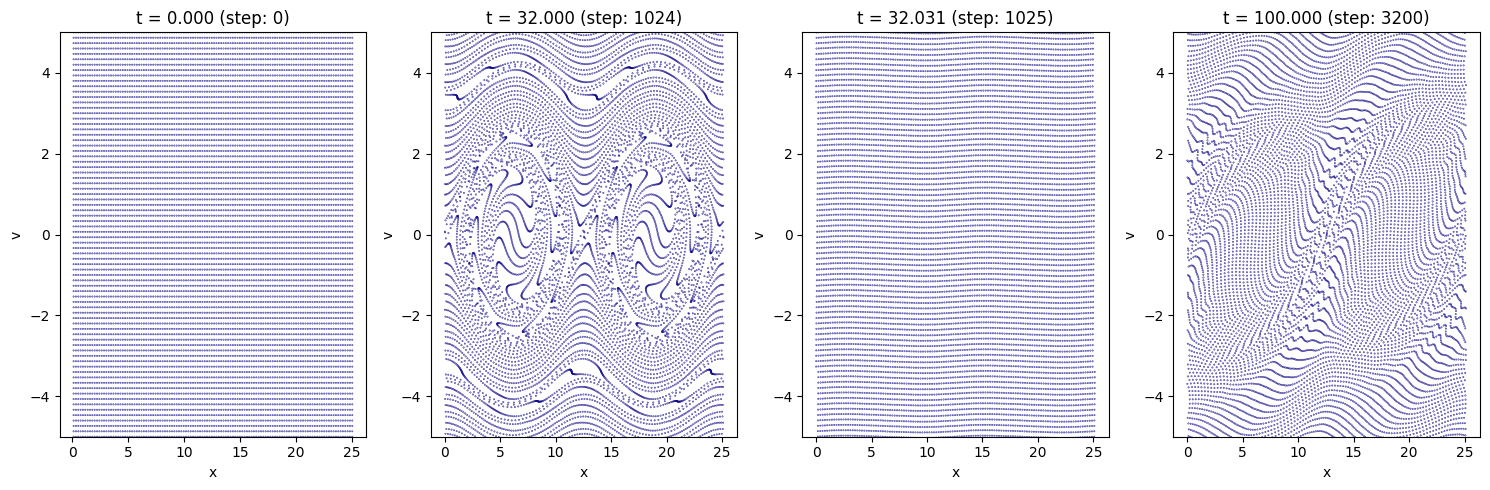}
    \caption{Two-stream instability phase-space particles.}
    \label{fig:TS_pp2}
\end{figure}

In Figure~\ref{fig:TS_pp2}, the particles are shown in phase-space with constant weights to remove coloring. For this case, a coarsened particle grid with $(60\times50)\times(2\times3)$ was used to better show the effect of resampling on grid preservation. As shown in the figure, prior to the resampling step, the phase-space is significantly distorted, with several regions underrepresented by particles or lacking particles altogether. After resampling, we return to a uniform, well-sampled grid. As shown in Figure~\ref{fig:TS_pp1}, over long enough time scales, this effect becomes more pronounced and the need for resampling is obvious. Large secondary vortices appear in phase-space at the top and bottom of the velocity space. In runs with no resampling or high resampling periods (top and middle rows of Figure~\ref{fig:TS_pp1}), these vortices are largely void of any particles, whereas, in lower resampling period runs (bottom row of Figure~\ref{fig:TS_pp1}), these vortices remain filled with particles and are thus well sampled. Thus, we have shown that the inclusion of the pseudoinverse resampling algorithm leads improved accuracy in the linear growth rate and a more consistently well-sampled grid at later simulation times. However, at early times, the effect of the resampling algorithm is minimal in this test. For a more thorough test of resampling methods, we turn to Landau damping.
}
\subsection{Nonlinear Landau damping}

\begin{figure}[h!]
    \centering
    \begin{subfigure}[b]{0.45\textwidth}
       \centering
       \includegraphics[width=1\linewidth]{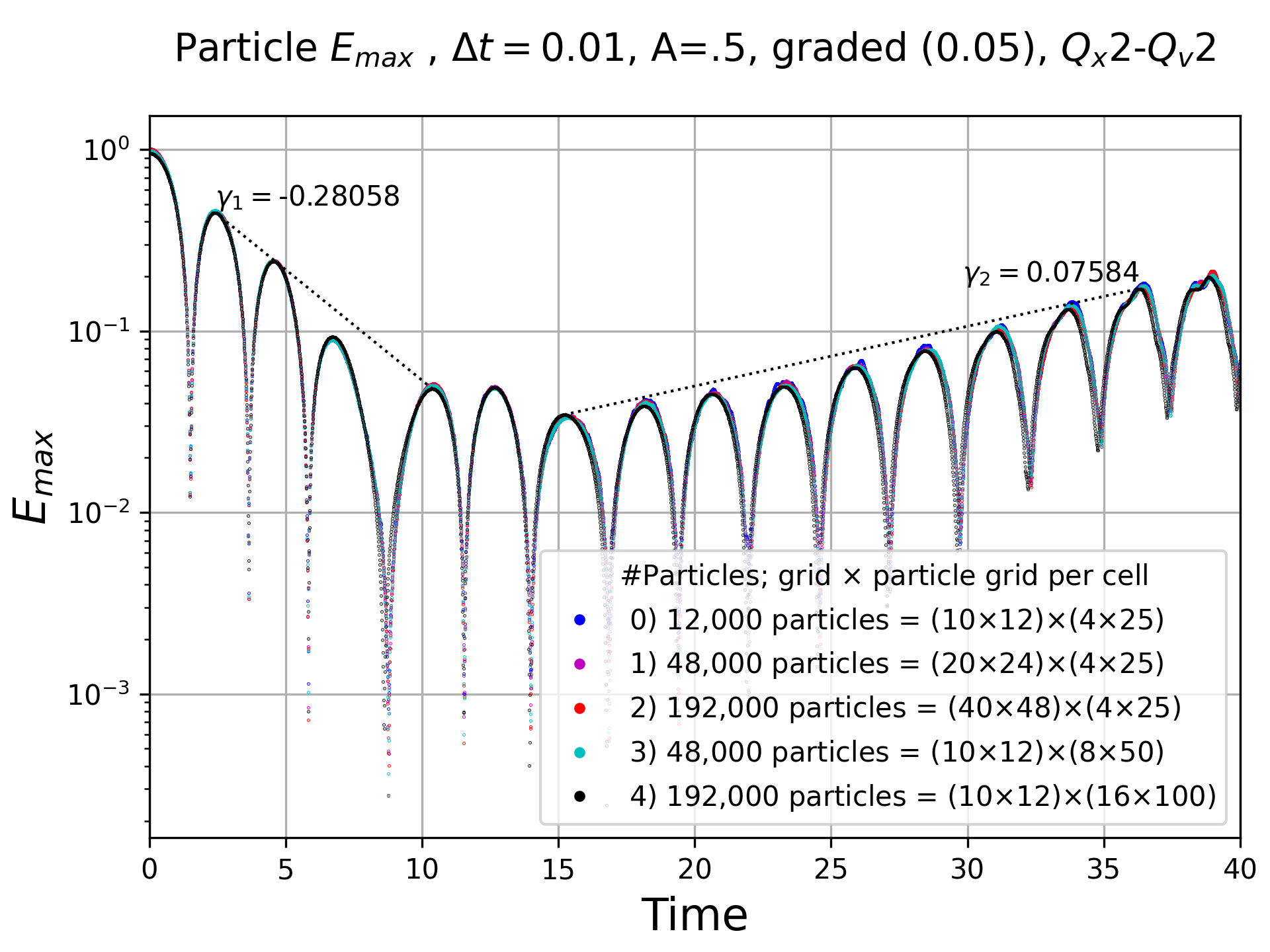}
       \caption{Particle number convergence}
    \end{subfigure}%
    \begin{subfigure}[b]{0.45\textwidth}
       \centering
       \includegraphics[width=1\linewidth]{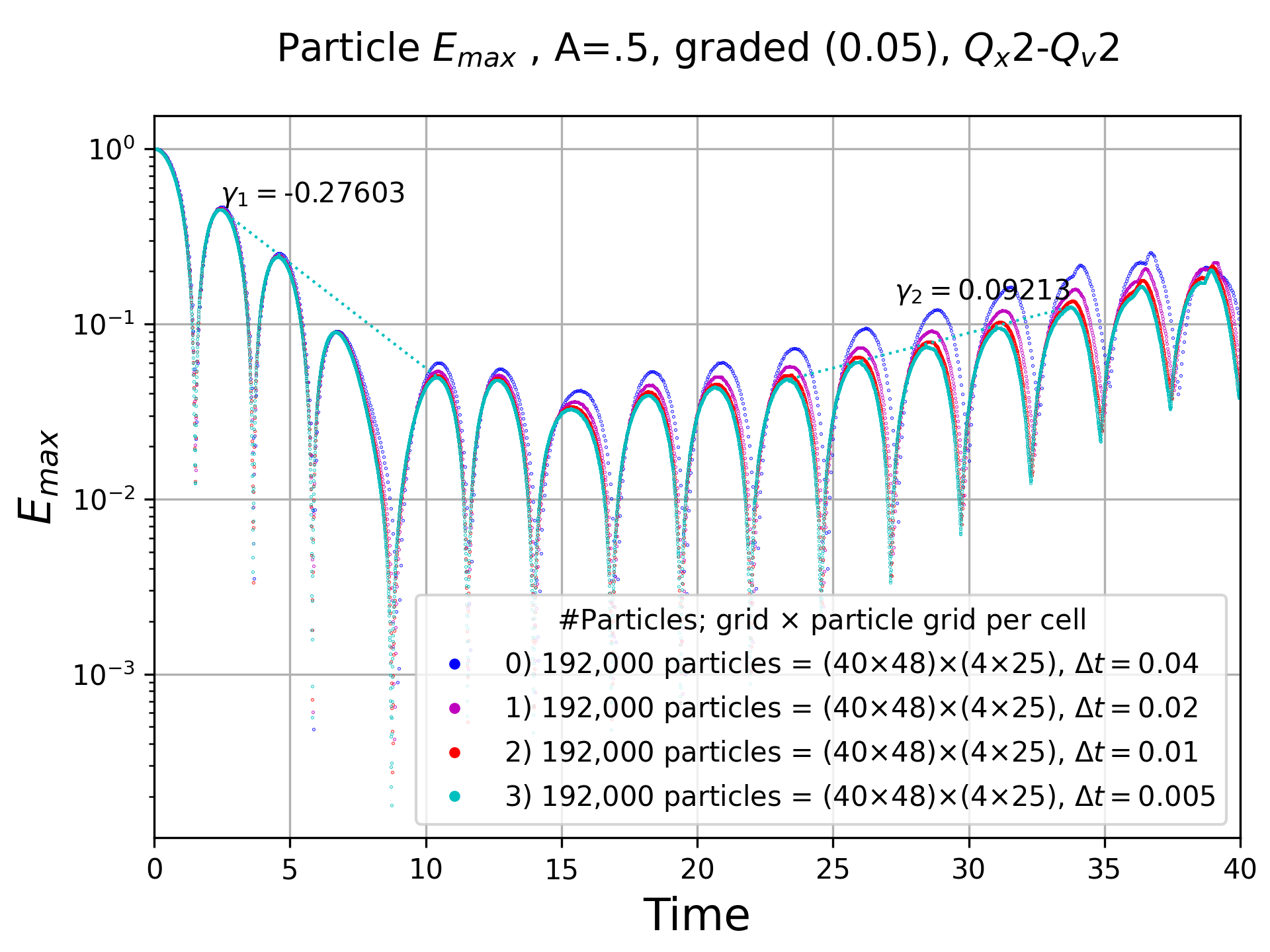}
       \caption{Convergence in $\Delta t$}
       \label{fig:A5_dt}
    \end{subfigure}%
     \caption{Converge study of nonlinear Landau damping, $A=0.5$}
    \label{fig:A5conv-1}
\end{figure}

\begin{figure}[h!]
    \centering    
    \begin{subfigure}[b]{0.45\textwidth}
       \centering 
       \includegraphics[width=1\linewidth]{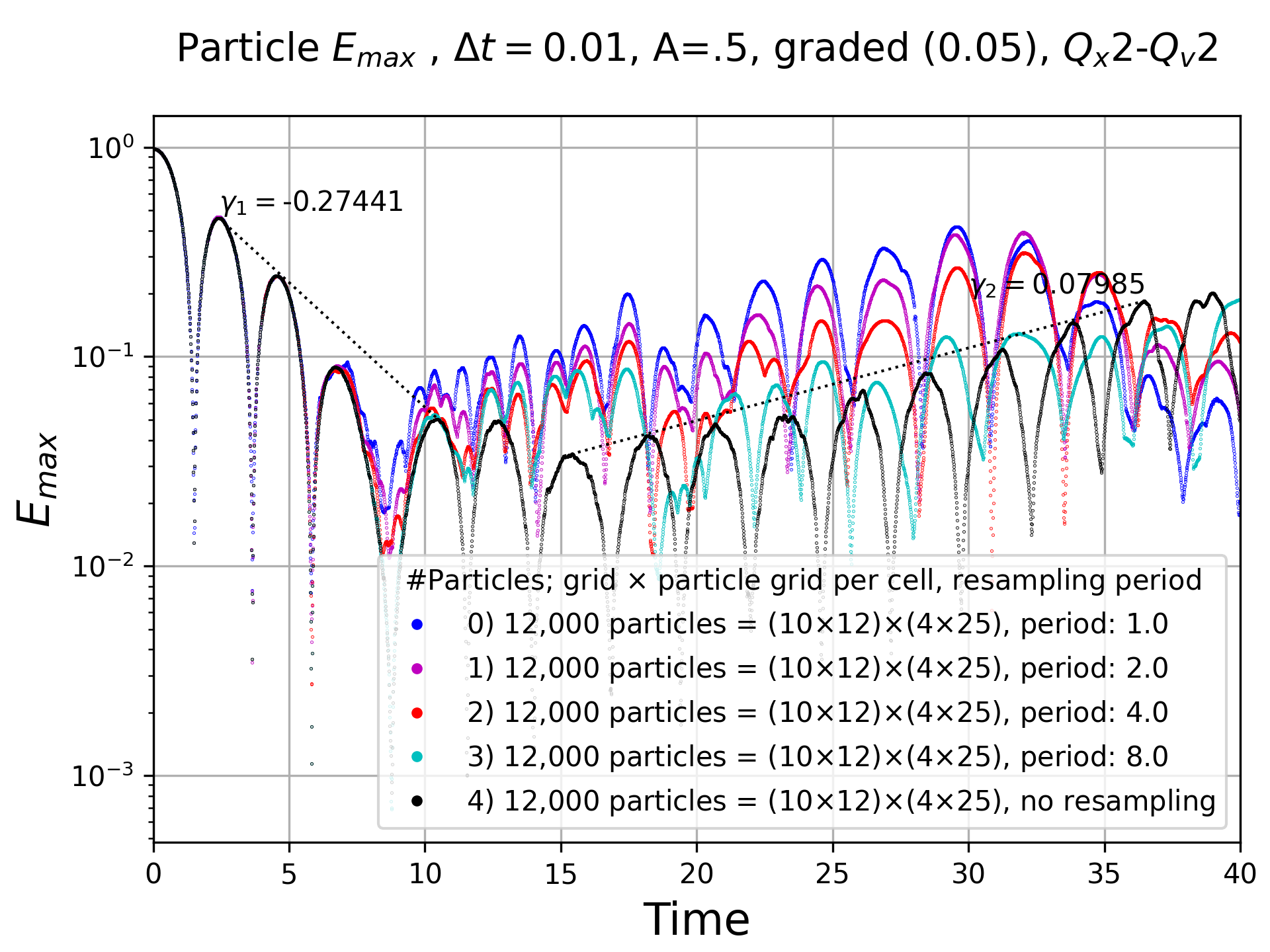}
       \caption{Convergence in resampling period}
       \label{fig:A5_period}
    \end{subfigure}%
    \begin{subfigure}[b]{0.45\textwidth}
       \centering 
       \includegraphics[width=1\linewidth]{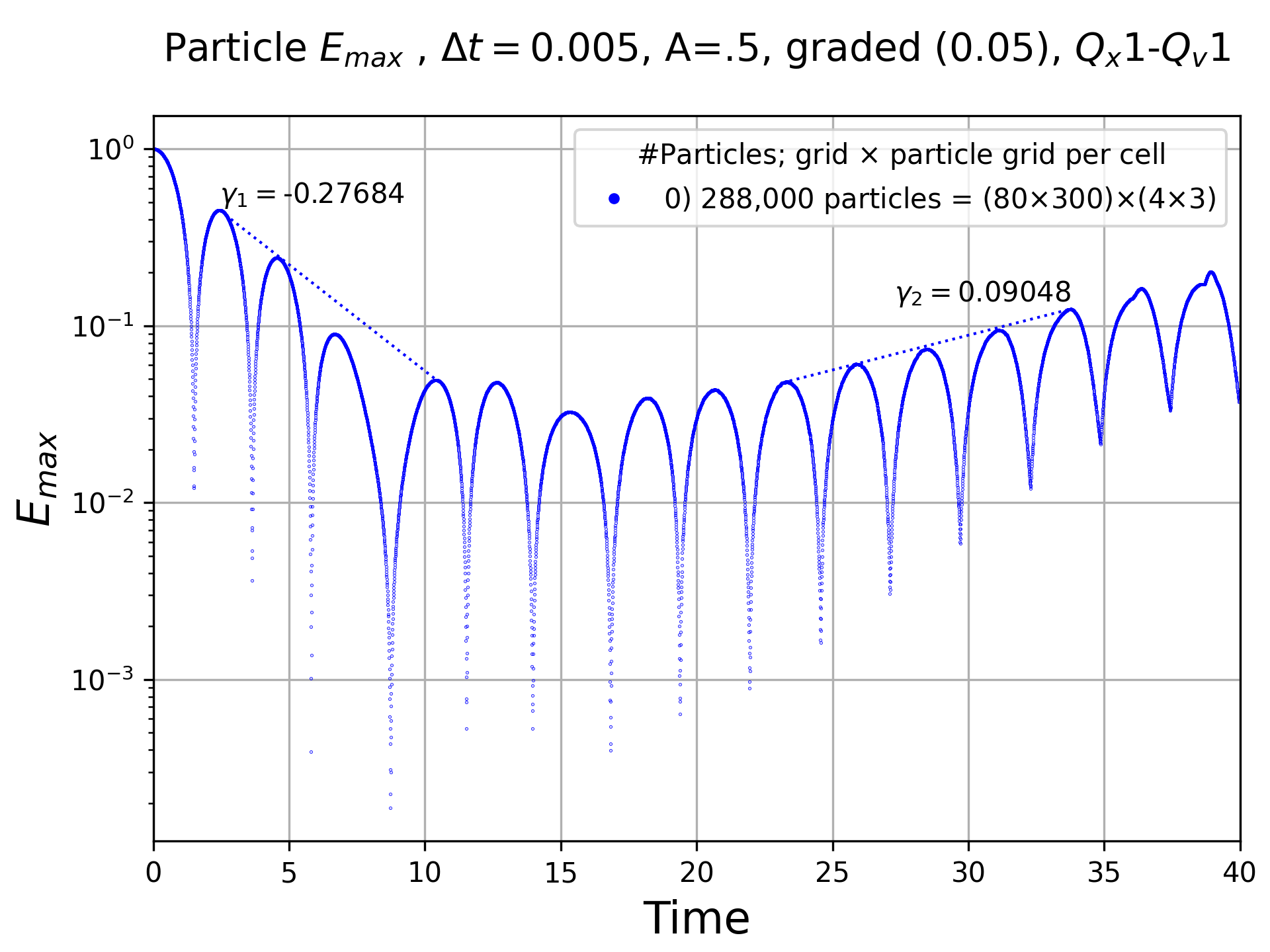}
       \caption{Highly converged}
       \label{fig:A5}
    \end{subfigure}%
    \caption{Convergence study of nonlinear Landau damping, $A=0.5$}
    \label{fig:A5conv-2}
\end{figure}

The nonlinear Landau damping test case, defined as setting $A=0.5$, has been studied extensively in the literature (Kraus Table 5.1 tabulates several results of previous work \cite{KrausThesis}).
Cheng and Knorr test with a $32 \times 128$ cell continuum grid solver and a time step of $\frac{1}{8}$ that is not only quantitatively similar to our results (Figure~\ref{fig:A5}) but qualitatively similar (Figure~4 in \cite{Cheng1976} and Figure~5 in \cite{Kraus_Kormann_Morrison_Sonnendrücker_2017} shows $\gamma_1$ measured with the first two peaks, and the fourth peak).
Figure~\ref{fig:A5conv-1} and \ref{fig:A5conv-2} show convergence studies on the amplitude of the electric field for time step, resampling rate and particle and continuum grid resolution.


These convergence tests establish parameters for an highly resolved analysis in Figure~\ref{fig:A5} that agrees well with Cheng and Knorr: $\gamma_1 = -0.281$ as compared to our rate of $-0.278$, and a growth rate of $0.084$ vs our rate of $0.090$.
Figure~\ref{fig:A5_dt} shows convergence study in time step where convergence is observed in the rebound region.
The modest effect of resampling in this nonlinear case is observed in Figure~\ref{fig:A5_period}, where resampling appears to increase the amplitude in the rebound stage.

\subsection{Linear Landau damping}
\label{sec:A0001}

\begin{figure}[h!]
    \centering
    \includegraphics[width=.8\linewidth]{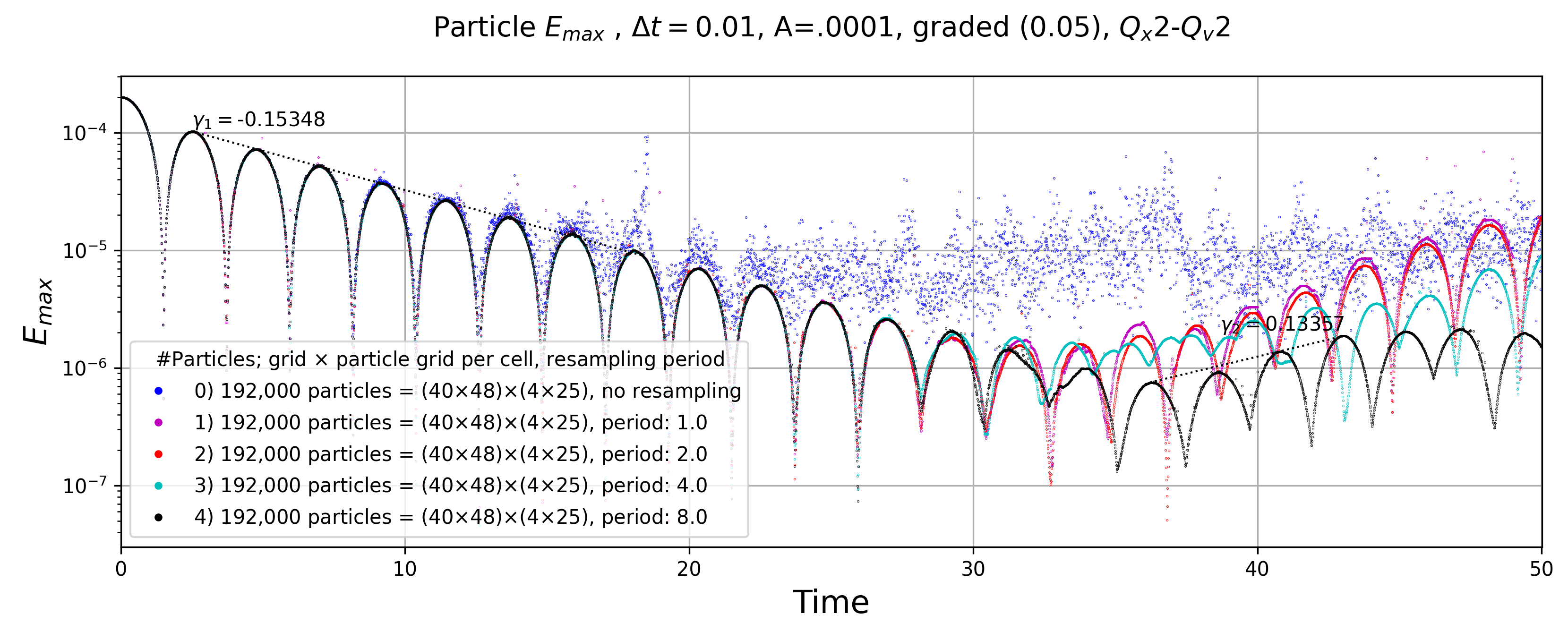}
    \caption{$E_{max}, A = 0.0001$, resample rate, and without resampling, ``convergence" test}
    \label{fig:A0001}
\end{figure}

The linear case of Landau damping, $A=0.0001$, is used here to demonstrate the potential of the projection algorithm.
Figure~\ref{fig:A0001} shows the electric field amplitude with a variety of resampling rates with $Q2$ spaces in both real space and velocity space. 
A damping rate of $\gamma_1 = -0.15348$ is observed, which agrees with theory reported in \cite{KrausThesis}.
With no resampling, the results are clearly very noisy while all of the tests with resampling, regardless of the rate, are free of noise. 
Resampling does, however, seem to kick the plasma into the growth phase faster, given that tests with a high resampling rate start to grow earlier. 

\section{\bt{Long time stability}}
\label{sec:longtime}
{\color{black}

Supporting long time, stable PIC simulation is a well known challenge for PIC methods and reducing noise, as we have demonstrated in even short time frames (Figures \ref{fig:myers} and \ref{fig:A0001}), is necessary to address this challenge.
This section investigates the behavior of our current method in long time simulations with Landau damping (\S\ref{sec:longtime1}) and two-stream (\S\ref{sec:longtime2}).

\subsection{Landau damping}
\label{sec:longtime1}

In an idealized, infinitely resolved grid, we expect the linear Landau damping problem to decay indefinitely~\cite{Villani2009}. In a discrete system, however, discretization errors create subgrid modes, with wavelengths bounded by the Nyquist mode, $k_{max} = \pi/\Delta x$. Thus, in coarser grids, high frequency velocity structures in the plasma may \textit{alias} onto lower-wavenumber subgrid modes, which will drive energy back into the field causing the ``rebound'' effect observed in essentially all collisionless experiments reported in the literature (Fig. 5.3 \cite{KrausThesis}). This effect can be suppressed by reducing the grid size, $\Delta x$, below the Debye length, $\lambda_D=\sqrt{1/n^3}$ in 1D normalized form, where $n=\int_{\Omega} f dv$ is the total plasma density. Previous work~\cite{Birdsall1980,Werner2025}, has suggested an ideal range for suppressing subgrid modes is $0.15\leq \lambda_D / \Delta_x \leq 1.0$. 
For our system, at the upper limit of this range ($\lambda_D=\Delta x$), given a density of $n=4\pi$ and a Debye length of $\lambda_D=0.022$, the number of grid cells will be 560 cells in space. 

Figure~\ref{fig:A0001-hi-res-ppc} shows a refinement study in the number of particles per cell which does not effect the location of the rebound. We further note that the amplitude of the rebound converges at about $1M$ particles as is shown in Figure~\ref{fig:A0001-hi-res-ppc-detail} and Figure~\ref{fig:A0001-hi-res-ppc} box ``(b)".
The amplitude of the second rebound (at time $T=100$ in units of inverse plasma frequency $[\omega_p^{-1}]$) is very close for all but the lowest particle count ($16K$) model and this alternating pattern slowly decays leaving coherent and noise-free waves (see detail at $T=500$ Figure~\ref{fig:A0001-hi-res-ppc-detail-end}).
However, the $16K$ particle model starts to heat up and continues to grow at longer times ($T=1,000$ data is available in the repository).
The $65K$ and $259K$ models remain stable with clean oscillations up to $T=1,000$.
The observed change in amplitude of the rebound may be explained by the plasma's increased ability to absorb wave energy as the particle density is increased. We expect this to converge monotonically but the additional presence of statistical noise in the system, as well as the rebound's apparent dependence on its state just before rebounding may offer an explanation as to why we do no observe a clean convergence structure. 

\begin{figure}[ht!]
       \centering
       \includegraphics[width=\linewidth]{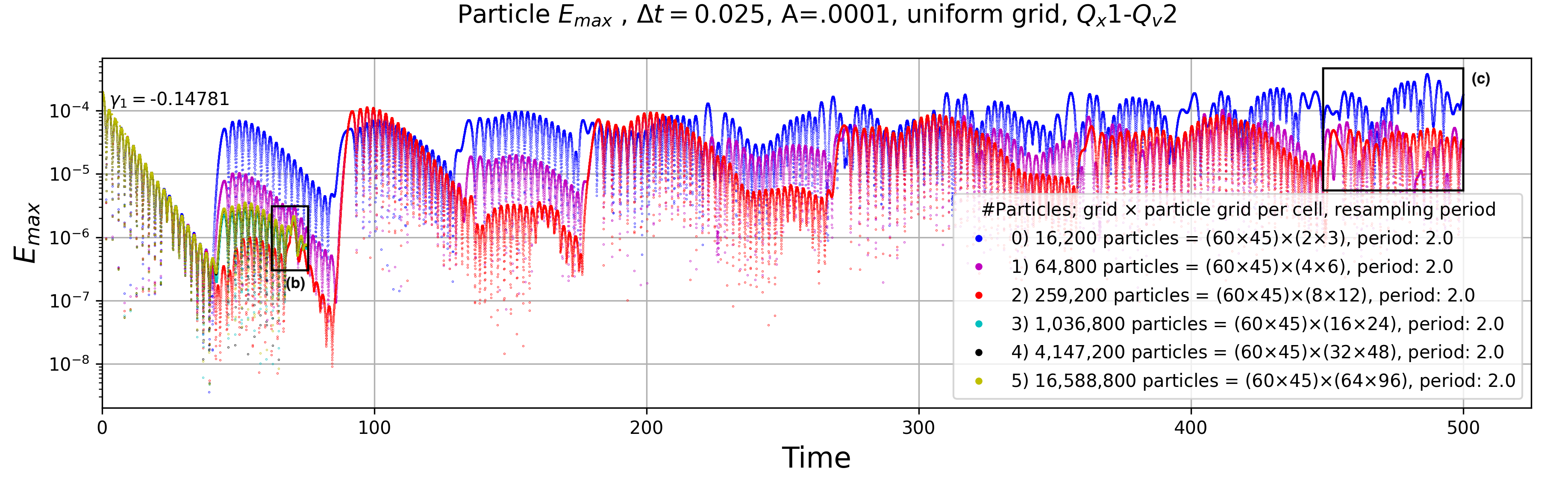}
       \caption{$E_{max}, A = 0.0001$, Refinement in particles per cell long time simulations}
       \label{fig:A0001-hi-res-ppc}
\end{figure}


Figure~\ref{fig:A0001-hi-res-ncells} shows refinement studies in grid density with constant number of particles per phase-space cell.  An increase in cell count delays the onset of subgrid modes and suppresses the rebound with the largest case almost entirely suppressed before $T=500$, when a large amplitude subgrid mode erupts and the continues with high-frequency, yet coherent, oscillations.
\begin{figure}[ht!]
    \centering
       \centering
       \includegraphics[width=\linewidth]{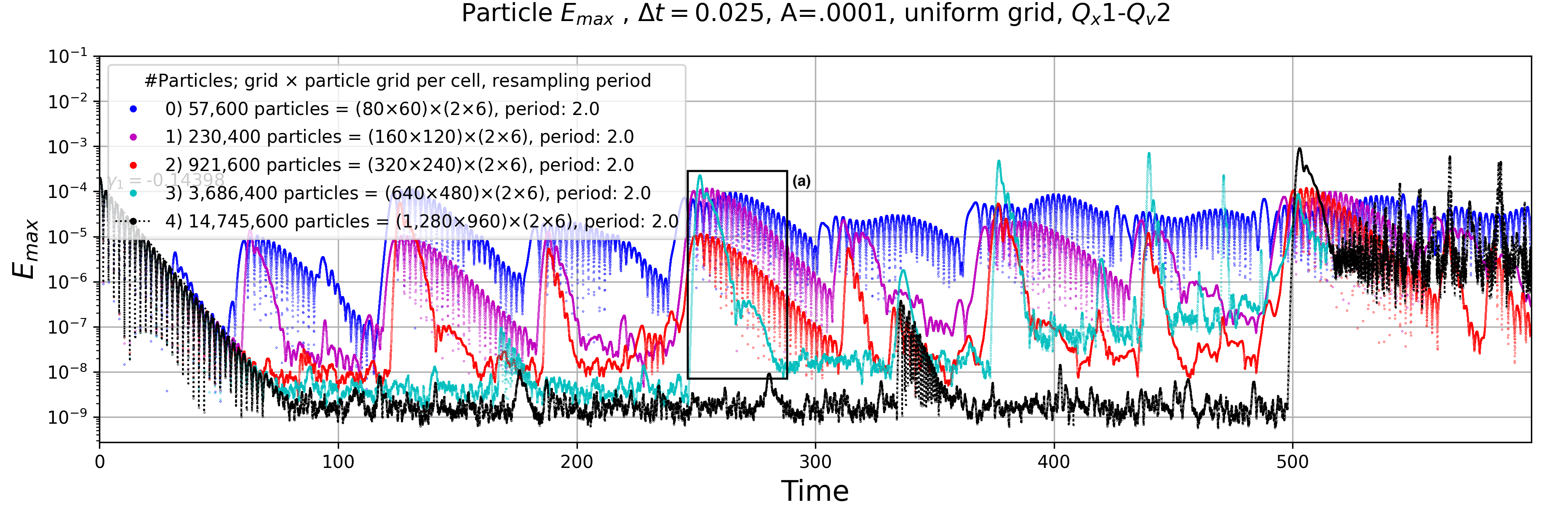}
       \caption{$E_{max}, A = 0.0001$, Refinement in grid density long time simulations}
       \label{fig:A0001-hi-res-ncells}
\end{figure} 
\begin{figure}[ht!]
    \begin{subfigure}[b]{.32\textwidth} \centering
       \centering
       \includegraphics[width=\linewidth]{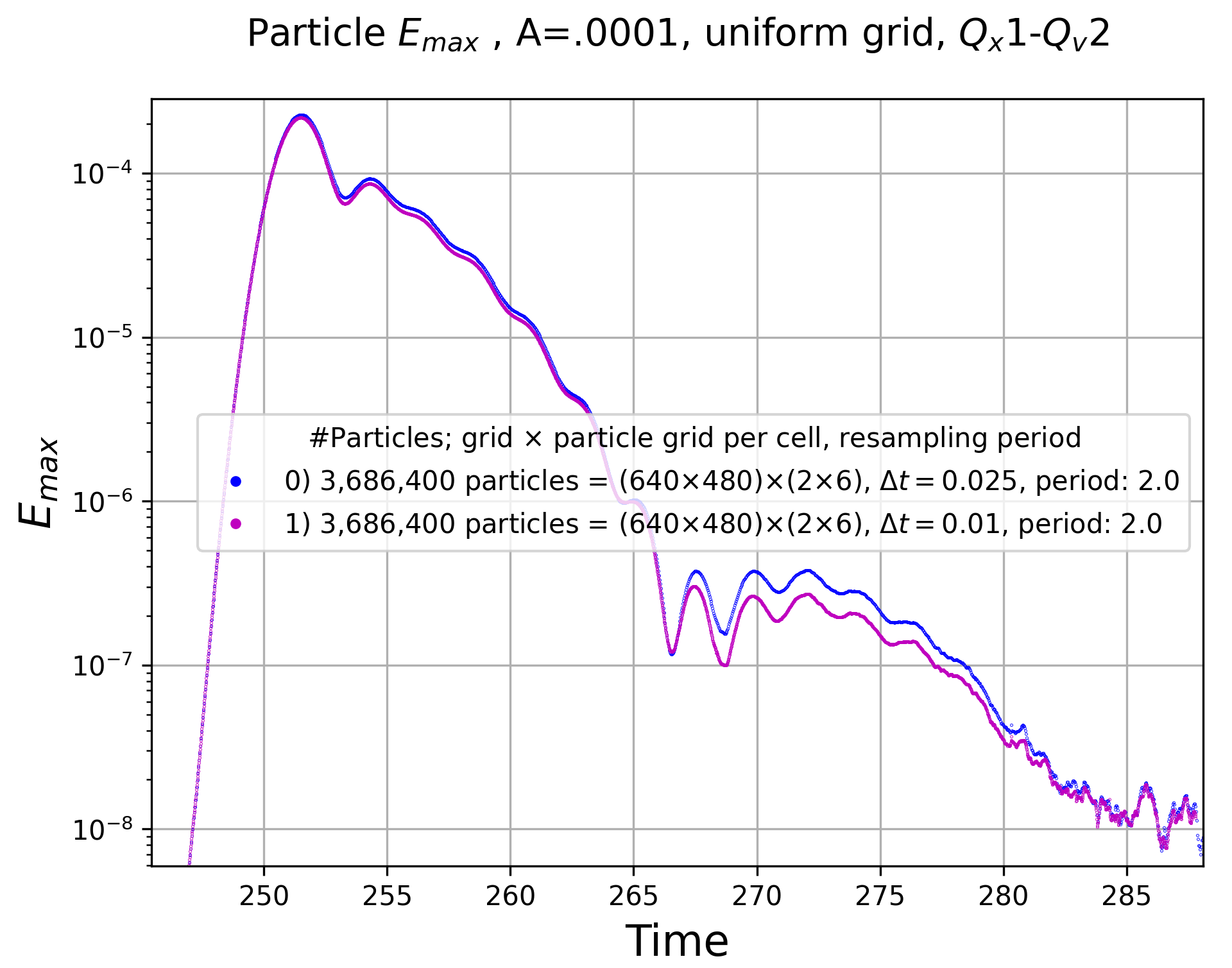}
       \caption{$\Delta t$ convergence ($\Delta t = 0.01$ and $\Delta t = 0.025$), $3.6M$ particles Fig.\ref{fig:A0001-hi-res-ncells} }
       \label{fig:A0001-hi-res-dt-conv}
    \end{subfigure}
    \begin{subfigure}[b]{.34\textwidth} \centering
       \centering
       \includegraphics[width=\linewidth]{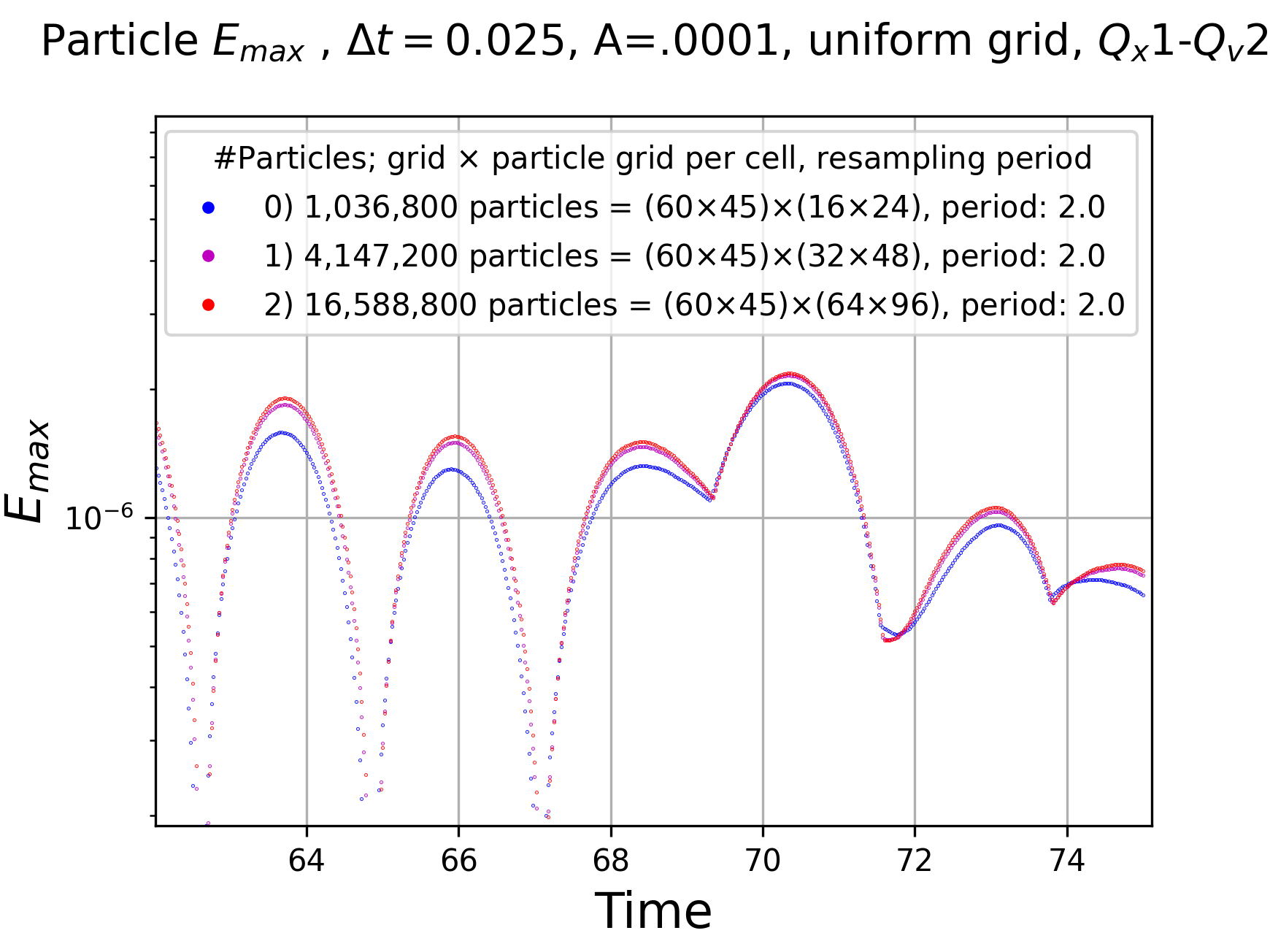}
       \caption{Convergence detail Fig. \ref{fig:A0001-hi-res-ppc} (box)}
       \label{fig:A0001-hi-res-ppc-detail}
    \end{subfigure}
    \begin{subfigure}[b]{.32\textwidth} \centering
       \centering
       \includegraphics[width=\linewidth]{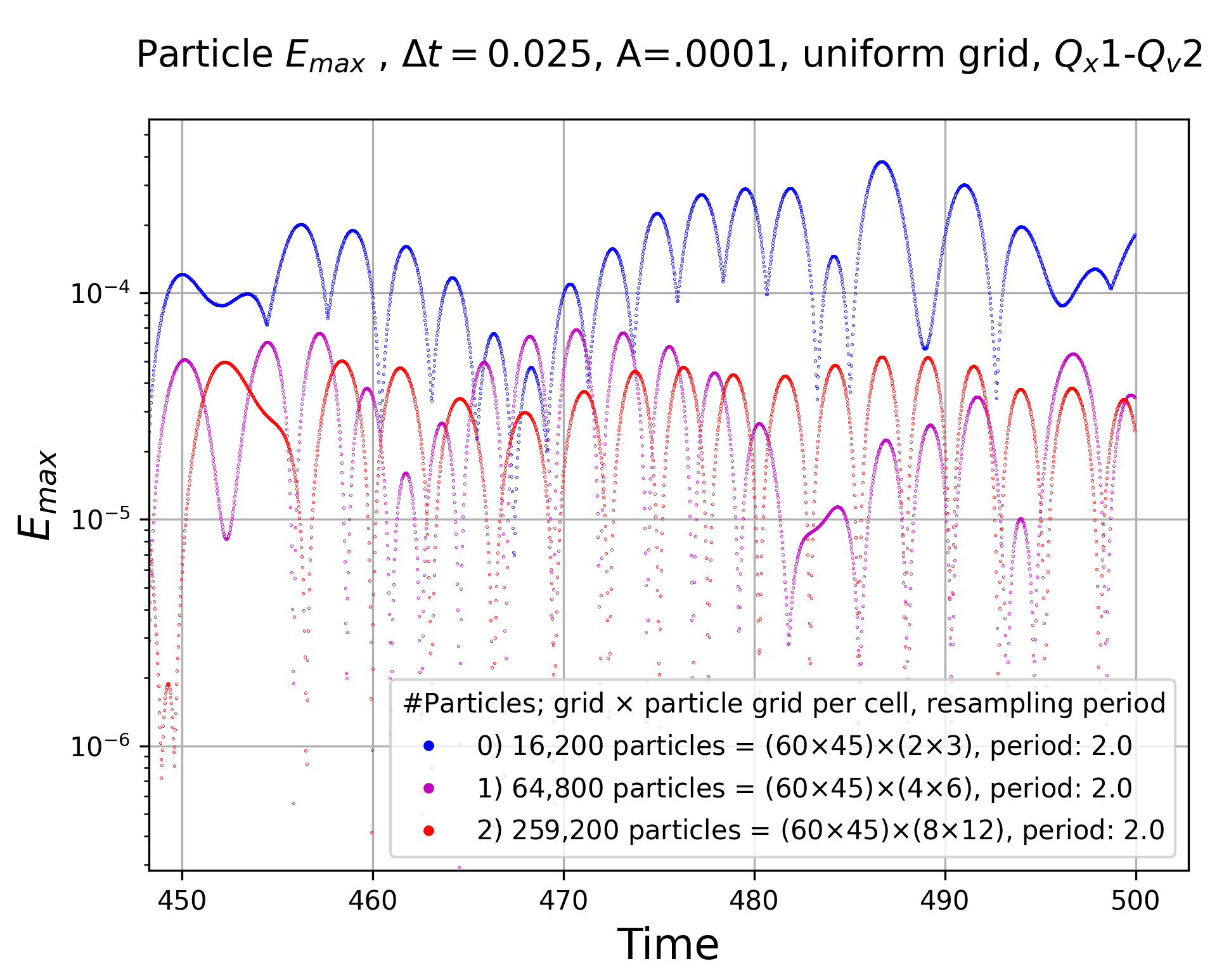}
       \caption{Long time stability Fig. \ref{fig:A0001-hi-res-ppc} (box)}
       \label{fig:A0001-hi-res-ppc-detail-end}
    \end{subfigure} \vspace{-.1in}
    \caption{$E_{max}, A = 0.0001$, long time simulations}
    \label{fig:A0001-hi-res}
\end{figure} 

The choice in timestep is based on previous works with PETSc-PIC~\cite{Finn2023,Finn2025} and Figure~\ref{fig:A0001-hi-res-dt-conv} shows little difference with a smaller time step than used in these studies. 



\pagebreak 

{\color{black}
\subsection{Two-stream}
\label{sec:longtime2}

As with the Landau damping tests, for long time stability, two-stream instability runs were extended to $T=500$ (Figure~\ref{fig:TS_ef12}, Figure~\ref{fig:TS_pp1}). After the initial growth phase, the simulation remains relatively stable through the entire nonlinear phase. Furthermore, we observe clear noise reduction in the nonlinear phase when applying the pseudoinverse resampling algorithm. In fact, while dynamics vary between runs, every run with resampling shows complete noise reduction throughout the full simulation. The ability of the resampled simulation to remain stable and preserve coherent phase-space structures over many plasma periods provides strong evidence that the PIC framework is correctly capturing both the linear and nonlinear dynamics of the two-stream instability and is not dominated by numerical noise, grid artifacts, or energy drift.

With resampling enabled, however, we observe small but regular oscillations in the electric field strength during the nonlinear phase and after the initial vortices merge. As shown in Fig.~\ref{fig:TS_pp1}, secondary vortices form and interact after $T=32$, trapping particles whose oscillatory motion in $x$ modulates the charge density and electric field strength. This may be a signature of nonlinear bounce dynamics, common in two-stream instability tests. Alternatively, these oscillations may arise from numerical instabilities, similar to those discussed in~\ref{sec:longtime1}. Fine-scale structures developed at later times will produce grid-scale oscillations which can alias onto the physical modes modulating the electric field. Both of these, physical and numerical, effects can appear similarly in the results and are therefore a topic for further study in future work.
}

\section{Side effects of resampling}
\label{sec:problems}

Figure~\ref{fig:A5_period} demonstrates that the dynamics of the plasma is not highly sensitive \red{to the resampling period} within a broad range (e.g., $1-8$ $[\omega_p^{-1}]$), but that rebound starts  earlier with higher resampling rates, indicating that resampling is disturbing the plasma to some extent.
\red{To investigate this effect}, Figure~\ref{fig:A0001-unstable} (left) shows a linear example with resampling every time step and increasing resampling periods.
The highest resampling rates clearly lead to instabilities and these instabilities demonstrate classic growing oscillations up to a small amplitude, see detail in Figure~\ref{fig:A0001-unstable} (right), that is growing in mean value and plateaus at about $1.0$.
Perhaps stability analysis could provide insight on this issue.
\begin{figure}[h!]
    \centering
    \includegraphics[width=.51\linewidth]{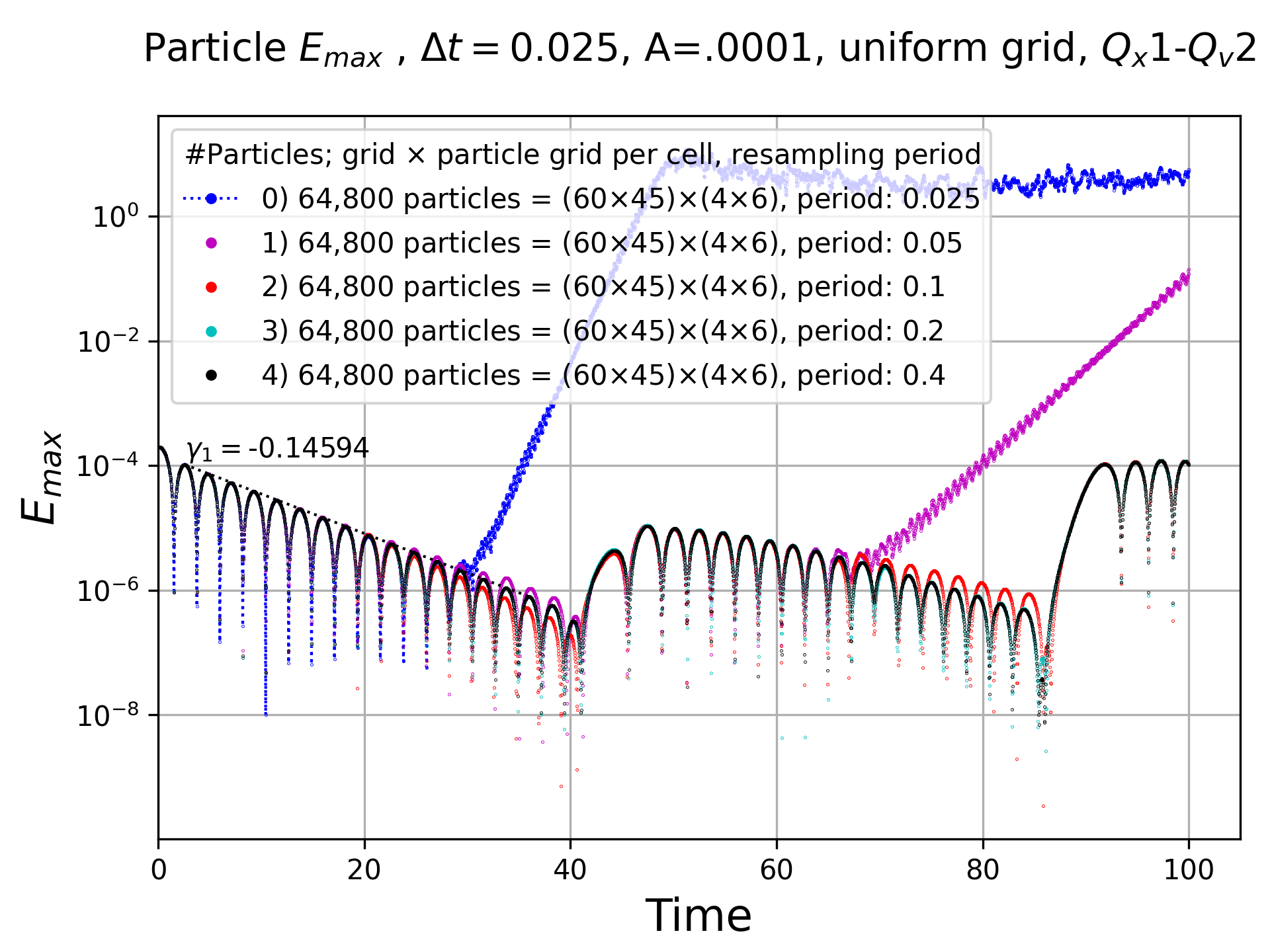}
    \includegraphics[width=.48\linewidth]{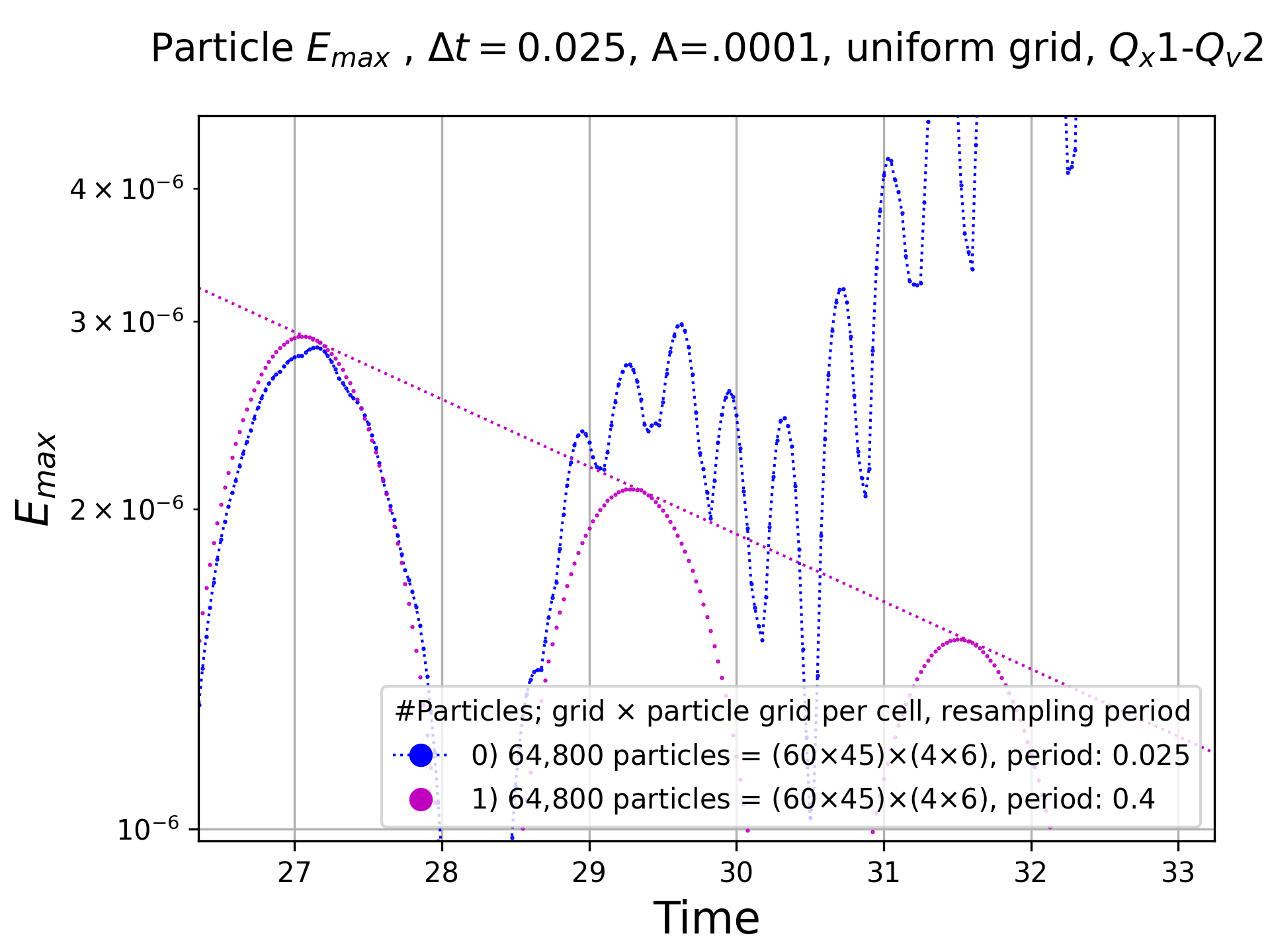}
    \caption{$E_{max}, A = 0.0001$, resampling period ``anti-convergence" study (left), instability detail (right)}
    \label{fig:A0001-unstable}
\end{figure}

While this instability at high resampling rates is not a practical problem, per se, in that all of our tests are stabilized with much lower resampling frequencies, this data does indicate that resampling can have adverse effects.
This test problem is highly idealized and does not provide convincing experimental evidence that the adaptivity strategy of
\red{simply mapping to the original grid} will be robust in practice.
More sophisticated adaptivity strategies are the subject of future work and would require more scientifically relevant models such as problems in magnetic reconnection in at least $2X + 3V$.
}

\subsection{\bt{Discussion}}
{\color{black}
Figure~\ref{fig:A0001-hi-res-ncells} sheds light onto the complex relationship and dependency between all the simulation variables, i.e., resampling rate, spatial and velocity grid density and time step. It is difficult to draw definitive conclusions about what exactly is driving the dynamics of the electric field damping and rebounding. As discussed in \S\ref{sec:longtime}, the grid resolution plays a large role in reducing the interaction of subgrid modes with the physical waves, while the particle density reduces stochastic, noise-driven heating effects in the plasma. These effects do not, however, converge monotonically suggesting that they are linked in some way. This requires further study in future work. 

The inclusion of collisions may also provide a smooth mechanism. The collision operator acts as a controlled, physically motivated diffusion term in the kinetic equations. This diffusive term continuously smooths out any build up of high-frequency components in the distribution function which can lead to spurious rebounds. In other algorithms (Fig. 5.5 \cite{KrausThesis}), clean decays to machine precision have been shown using collisions. Collisions must be used with some level of caution, however. Much of the essential collisionless interactions in the Landau damping test, in particular in the nonlinear regime, rely on fine-scale interactions between the particles and waves. 
Introducing a collision operator can have the effect of smoothing out those essential fine-scale interactions which, if too strong, will disrupt the fundamental physics of the system.
Recent work in the PETSc library~\cite{AdamsHirvijokiKnepleyBrownIsaacMills2017,pusztay2023landau,Adams2022a,adams2024performance} has provided a variety of collision algorithms that can be included in the PETSc-PIC framework in future work.
}
\section{Conclusion}
\label{sec:conc}

The paper develops a new approach to particle resampling that uses a conservative projection, a pseudoinverse, to map any distribution of particles to essentially any other distribution of particle while conserving all moments up to the degree of polynomial that the projection function space can represent exactly.
This method is evaluated with a static particle and continuum Cartesian grids, simply remapping to the original particle grid, on standard \red{two-stream instability and Landau damping test problems.
Two-stream instability tests show minor accuracy improvements when using resampling methods, as well as a well maintained particle grid over long time simulations. In Landau damping} tests, where noise is problematic, the linear cases, show that resampling reduces noise considerably and coherent dynamics are maintained for long times, where as the solution becomes essentially all noise without resampling.

This work suggests several areas of future work, namely:
\paragraph{Adaptivity strategies} Developing strategies to optimize the particle grids with respect to minimizing disturbances to the physics from resampling.
\paragraph{Entropy} With entropy measures, from our particle Landau collision operator \cite{pusztay2023landau}, we can determine the continuum grids required for resampling or a continuum collision operator \cite{adams2024performance,pusztay2023landau} to keep entropy generation by the projection well below the entropy generated by the physics.
\paragraph{Continuum grid AMR} AMR in velocity and regular particle grids on each cell of an adapted grid, like a cubed sphere \cite{adams2024performance}, is a path for generating adapted particle grids through continuum mesh adaptation.
\paragraph{Splitting and coalescing} These ideas, developed in many particle resampling methods, would allow for an incremental modification of the particle mesh to minimize cost and perhaps have less impact the dynamics.
\paragraph{Increasing relevance} Understanding the effects of resampling on physics, beyond conserving moments and other structure like entropy stability, requires experimentation with more complex models such as the Ion Temperature Gradient (ITG) instability to understand the robustness of these ideas. 

\section*{Acknowledgments}

This work was supported by the U.S. Department of Energy,
Office of Science, Office of Fusion Energy Sciences and Office of Advanced Scientific Computing Research,
Scientific Discovery through Advanced Computing (SciDAC) Partnership programs and the FASTMath Institute under Contract No. DE-AC02-05CH11231 at Lawrence Berkeley National Laboratory, the Office of Naval Research and by an appointment to the NRC Research Associateship Program at the U.S. Naval Research Laboratory, administered by the Fellowships Office of the National Academies of Sciences, Engineering, and Medicine. DSF and MGK were partially supported by NSF CSSI grant 1931524.

\appendix

\section{Artifact description and reproducibility}
\label{sec:ad}

PETSc output files with all data, provenance information, and reproducibility instructions for all tables and plots can be obtained from \path{git@gitlab.com:markadams4/resampling-paper.git}.
This includes the python scripts that generates the plots and run scripts, makefiles and PETSc resource files used to generate the data, and the test harness code in {\path{src}.
The \path{src/A.X} directories has data for $A=0.$X.
The exact PETSc versions (SHA1) are in the data files, with the provenance data, all parameters used in each test, but any PETSc version from v3.22 should suffice to reproduce this data. 

\bibliographystyle{siamplain}
\bibliography{mark_sp}

\end{document}